\documentclass[twocolumn,pre,showpacs,eqsecnum,floatfix]{revtex4}

\usepackage{dcolumn}
\usepackage{amsfonts}
\usepackage{amsmath}
\usepackage{amssymb}
\usepackage{bm}

\DeclareMathOperator{\tr}{tr}

\newif\ifpdf
\ifx\pdfoutput\undefined
\pdffalse 
\else
\pdfoutput=1 
\pdftrue \fi

\ifpdf
\usepackage[pdftex]{graphicx}
\else
\usepackage{graphicx}
\fi

\begin{document}

\ifpdf \DeclareGraphicsExtensions{.jpg,.pdf,.tif} \else
\DeclareGraphicsExtensions{.eps,.jpg} \fi

\setlength{\arraycolsep}{0.7mm}

\newcommand{\brm}[1]{\bm{{\rm #1}}}
\newcommand{\tens}[1]{\underline{\underline{#1}}}

\title{Anomalous elasticity of nematic and critically soft elastomers}

\author{Olaf Stenull}
\affiliation{
Department of Physics and Astronomy\\
University of Pennsylvania\\
Philadelphia, PA 19104\\
USA }

\author{T.~C.~Lubensky}
\affiliation{
Department of Physics and Astronomy\\
University of Pennsylvania\\
Philadelphia, PA 19104\\
USA }

\vspace{10mm}
\date{\today}

\begin{abstract}
\noindent Uniaxial elastomers are characterized by five elastic
constants.  If their elastic modulus $C_5$ describing the energy
of shear strains in planes containing the anisotropy axis
vanishes, they are said to be soft. In spatial dimensions $d$ less
than or equal to $3$, soft elastomers exhibit anomalous elasticity
with certain length-scale dependent bending moduli that diverge
and shear moduli that vanish at large length-scales. Using
renormalized field theory at $d=3$ and to first order in
$\varepsilon=3-d$, we calculate critical exponents and other
properties characterizing the anomalous elasticity of two soft
systems: (i) nematic elastomers in which softness is a
manifestation of a Goldstone mode induced by the spontaneous
symmetry breaking associated with a transition from an isotropic
state to a nematic state and (ii) a particular version of what we
call a critically soft elastomer in which $C_5=0$ corresponds to a
critical point terminating the stability regime of a uniaxial
elastomer with $C_5>0$.
\end{abstract}


\maketitle

\section{Introduction}

\noindent Liquid-crystalline
elastomers~\cite{FinKoc81,deGennes,Terentjev99} are elastic media
with the macroscopic symmetry properties of liquid
crystals~\cite{deGennesProst93,Chandrasekhar92}. They consist of
weakly crosslinked polymeric networks with mesogenic units. The
existence of the rubbery crosslinked network has apparently little
impact on liquid crystalline phase behavior. In fact, the usual
thermotropic liquid crystal phases, i.e., the nematic,
cholesteric, smectic-$A$, and smectic-$C$ phases have their
elastomeric counterparts~\cite{Terentjev99,KimFin01}. However,
because liquid-crystalline elastomers cannot flow, they have
mechanical properties that differ significantly from standard
liquid crystals. Usually, liquid-crystalline elastomers are
prepared by crosslinking side-chain~\cite{WarTer96} or
main-chain~\cite{Zentel89} polymers. Alternative methods of
synthesis include the polymerization of monomeric solutes in a
liquid-crystalline solvent~\cite{Bowman97} or the confinement of a
conventional liquid crystal in a dilute flexible matrix such as
aerosil~\cite{CrawfordZumer96}.

The main subject of this paper are nematic liquid-crystalline
elastomers, or briefly, nematic elastomers (NEs). For recent
reviews on NEs see Refs.~\cite{WarTer96,Lubensky&Co_2001}. These
materials have unique properties that make them candidates for
device applications. Temperature
change~\cite{kuepfer_finkelmann_94} or
illumination~\cite{finkelmann_etal_01} can alter the orientational
order and cause  the elastomer to extend or contract as much as
$400\%$~\cite{FinWer00}. This qualifies nematic elastomers as
contestants for use in artificial muscles~\cite{hebert,ratna}.
Another striking property of nematic elastomers is their soft
elasticity~\cite{golubovic_lubensky_89,FinKun97,VerWar96,War99}
characterized by vanishing shear stresses for a range of
longitudinal strains applied perpendicular to the uniaxial
direction.

The origin of soft elasticity in NEs is the spontaneous breaking
of rotational symmetry of the isotropic state induced by the
development of orientational order in the nematic
state~\cite{golubovic_lubensky_89,Olmsted94}. This spontaneous
symmetry breaking has the consequence that NEs are not like
conventional uniaxial elastomers characterized by five independent
shear moduli. Rather, the elastic constant associated with shear
in planes containing the anisotropy axis (customarily called
$C_5$) vanishes in NEs.

Though we are primarily interested in NEs, we also study a
particular version of another class of soft uniaxial elastomers in
which the elastic constant $C_5$ simply vanishes for energetic or
entropic reasons.  In this case, $C_5=0$ is a critical point
marking the boundary between the high-symmetry uniaxial phase with
$C_5>0$ and a low-symmetry sheared phase with $C_5 <0$.  In other
words, $C_5$ acts like the temperature variable in a standard
thermal phase transition.  A complete model of this critical point
requires the introduction of third and fourth order terms in the
nonlinear strains to stabilize the system when $C_5<0$.  Since
this complete model is characterized by a large number of
parameters, it is quite complex.  Rather than analyze this full
model, we consider simpler model systems, which we call critically
soft elastomers or CSEs, defined by simply setting $C_5=0$ in the
standard elastic energy of a uniaxial medium containing only
quadratic terms in nonlinear strains.  Remarkably, CSEs exhibit
well-defined anomalous elasticity much like that of the more
physical NEs even though they lack the nonlinear terms needed to
stabilize their low-symmetry phase. CSEs are simpler in many ways
than NEs, and the analysis of their anomalous behavior provides a
useful and instructive tutorial prelude to the analysis of NEs.

On the level of mean-field theory, the elastic energies of NEs and
CSEs coincide. Therefore, their respective elastic properties are
equivalent above their mutual upper critical dimension 3. For
dimensions $d \leq 3$, however, fluctuations become important.
These fluctuations lead to
Grinstein-Pelcovits~\cite{grinstein_pelcovits_81_82} type
renormalizations culminating in anomalous elasticity, i.e., in a
length-scale dependence of certain elastic constants, with
different universality classes for NEs and CSEs.

In the present paper we explore the anomalous elasticity of NEs
and CSEs by carrying out a renormalization group (RG) analysis.
Using the methods of renormalized field
theory~\cite{amit_zinn-justin} we examine the scaling behavior of
the elastic constants in $d=3$ as well as in $d=3-\varepsilon$
dimensions. A brief account of our work on NEs appears in
Ref.~\cite{stenull_lubensky_epl2003}. Our work on CSEs has not
been reported hitherto. We will treat only systems in which random
stresses are not important.  Random stresses lead to a different
universality class with anomalous
behavior~\cite{golubovic_lubensky_89,Xing_Radz_03} below five
rather than three dimensions.

The plan of presentation is as follows: In Sec.~\ref{resSum} we
give a brief summary of our main results. In Sec.~\ref{ise} we
discuss CSEs. In Sec~\ref{iseModel} we briefly review some
elements of the Lagrangian theory of elasticity and then set up a
Landau-Ginzburg-Wilson elastic energy functional (Hamiltonian) for
CSEs. As a prelude to the subsequent RG analysis we analyze the
symmetry contents of this Hamiltonian. Section~\ref{iserg}
contains the core of our renormalized field theory of CSEs. We
explain our diagrammatic perturbation calculation and its
renormalization. By solving the appropriate RG equation, we
ascertain the scaling behavior of displacement correlation
functions and ultimately that of the relevant elastic moduli. We
conclude Sec.~\ref{ise} by making contact with conventional
uniaxial elastomers by incorporating a small but nonvanishing
$C_5$, which leads to semi-soft elasticity. Section~\ref{ne} deals
with NEs and has an outline similar to that of Sec.~\ref{ise}. In
Sec.~\ref{neModel} we derive a Landau-Ginzburg-Wilson minimal
model for NEs in form of a field theoretic Hamiltonian. Our
renormalized field theory for this model is presented in
Sec.~\ref{nerg}. The main part of this paper concludes with
Sec.~\ref{summary}, where we give  concluding remarks. There are
five Appendices. In Appendices~\ref{app:ward1}, \ref{app:ward2}
and \ref{app:ward3} we derive Ward identities for CSEs and NEs. An alternative RG approach to NEs is sketched in Appendix~\ref{app:alternativeScheme}. Appendix~\ref{app:diagramCalc} contains details on the calculation
of Feynman diagrams.

\section{Summary of results}
\label{resSum}
For the convenience of the reader we now summarize our main
results before we get into details of our work.

\subsection{Critically soft elastomers}
The elastic constants of CSEs are defined via the model elastic
energy
\begin{eqnarray}
\label{CSEdefEn}
&&\mathcal{H} = \frac{1}{2} \int d^{d_\perp} x_\perp \int d x_d \,
\Big\{  C_1 \, u_{dd}^2 + K \left(  \nabla_\perp^2 u_d \right)^2
 \\
&&   + \, 2 C_2 \,  u_{dd} u_{aa}    +  C_3 \, u_{aa}^2 + 2C_4 \,
u_{ab} u_{ab} + C_5 \, u_{ad} u_{ad} \Big\} \, , \nonumber
\end{eqnarray}
where $u_d$  and $u_a$, $a=1, ... , d-1$, are, respectively, the
directions parallel and perpendicular to the nematic order and
$u_{dd}$, $u_{ab}$, and $u_{ad}$ are components of the Lagrangian
nonlinear strain
tensor~\cite{landau_lifshitz_elasticity,tomsBook}. The elastic
constant $C_1$ describes longitudinal shear along the anisotropy.
$C_2$ couples strains along the anisotropy axis to shears in the
plane perpendicular to the anisotropy axis. The elastic constants
$C_3$ and $C_4$ are associated with shear purely in the plane
perpendicular to the distinguished direction. $K$ is a bending
modulus. The pure CSE system with soft elasticity is characterized
by $C_5=0$. The coupling $C_5$  plays a role similar to that of
the temperature in a thermal phase transition. When $C_5>0$, the
system displays conventional uniaxial elasticity. When $C_5<0$,
the system is unstable with respect to the formation of a
lower-symmetry (sheared) elastic state. $C_5=0$ marks the
transition between the two phases. At this point, the system shows
critical behavior analogous to critical behavior at a thermal
phase transition. Our simple CSE model does not include higher
order terms in the strain to stabilize the sheared phase. We
expect, however, that the full model with these terms included
will have the same structure as our CSE model, at least for $C_5
\geq 0$.

The inverse displacement correlation functions (vertex functions)
obey scaling forms that can be expressed for $d<3$ as
\begin{subequations}
\label{CSEscaleForms}
\begin{eqnarray}
\Gamma_{dd}(\brm{q}_{\perp} , q_d) &= &\frac{K}{T} L_{\perp}^{-4}
\ell^{4- \eta_K}
 \\
&\times&  {\hat \Phi}_{dd} \left(
\frac{L_{\perp}\brm{q}_{\perp}}{\ell^{1}} , \frac{L_d
q_d}{\ell^{\phi}}, \frac{C_5}{K} \frac{L_{\perp}^2}{\ell^{1/\nu_5
}}  \right) \, ,
\nonumber \\
\Gamma_{ad}(\brm{q}_{\perp} , q_d) &= &\frac{C_2}{T}
L_{\perp}^{-1} L_d^{-1} \ell^{1+ \eta_2 + \phi}
 \\
&\times&  {\hat \Phi}_{ad} \left(
\frac{L_{\perp}\brm{q}_{\perp}}{\ell^{1}} , \frac{L_d
q_d}{\ell^{\phi}}, \frac{C_5}{K} \frac{L_{\perp}^2}{\ell^{1/\nu_5
}}  \right) \, ,
\nonumber \\
\Gamma_{ab}(\brm{q}_{\perp} , q_d)&=&
\frac{C_4}{T}L_{\perp}^{-2}\ell^{2+\eta_C}
\\
&\times& {\hat \Phi}_{ab}
 \left( \frac{L_{\perp}\brm{q}_{\perp}}{\ell^{1}} , \frac{L_d q_d}{\ell^{\phi}}, \frac{C_5}{K} \frac{L_{\perp}^2}{\ell^{1/\nu_5 }}  \right) \, ,
\nonumber
\end{eqnarray}
\end{subequations}
where $q_d$ and $\brm{q}_{\perp}$ are, respectively, wavenumbers
(or momenta) parallel and perpendicular to the anisotropy axis.
\begin{subequations}
\begin{eqnarray}
\eta_K = 4 \varepsilon/7 \, ,\quad \eta_C = \varepsilon/7 \, ,
\quad \eta_2 =  \varepsilon/7 \, ,
\\
\nu_5 = 1/2 - \varepsilon/28 \, , \quad \phi = (4 - \eta_K)/2 \, ,
\end{eqnarray}
\end{subequations}
where $\varepsilon = 3-d$, are scaling exponents. In principle,
the exponents $\phi$ and $\eta_2$ could be independent of the
other exponents. For the CSE model, however, they are not at least
to first order in $\varepsilon$. The scaling functions ${\hat
\Phi}_{dd}(\brm{q}_{\perp}, q_d, e)$, ${\hat
\Phi}_{ad}(\brm{q}_{\perp}, q_d, e)$, and
${\hat\Phi}_{ab}(\brm{q}_{\perp}, q_d, e)$ are respectively
proportional to $\brm{q}_{\perp}^4 + q_d^2 + e \,
\brm{q}_\perp^2$, $\brm{q}_\perp q_d$ and $\brm{q}_{\perp}^2$ in
the long-wavelength limit in mean-field theory. ${\hat\Phi}_{ab}$
also has a term proportional to $q_d^4$, but its coefficient is
irrelevant and we will not be concerned with it here. The boundary
between scaling and Gaussian behavior is marked by the non-linear
length scales
\begin{subequations}
\begin{eqnarray}
L_\perp &\sim& \left[ \sqrt{C_1 K^3} /(C_4 T)
\right]^{1/\varepsilon}  ,
\\
L_d &=& \sqrt{C_1/K} \, L_\perp^2 \, ,
\end{eqnarray}
\end{subequations}
where $T$ is the temperature measured in units so that the
Boltzmann constant is equal to 1.

The above scaling forms imply the following scaling for the
elastic constants:
\begin{eqnarray}
\label{revCSEC}
C_2 &\sim& C_3 \sim C_4 \\
&\sim&  \left\{
\begin{array}{lcl}
(L_\perp |\brm{q}_\perp|)^{\eta_C} &\ \mbox{if} \ & \ \xi_5
|\brm{q}_\perp| \gg 1\, , \ q_d =0
\\
(L_d q_d)^{\eta_C/\phi} &\ \mbox{if} \ & \ \xi_5  |\brm{q}_\perp|
L_d /L_\perp\gg 1 \, , \ \brm{q}_\perp = \brm{0}
\\
(L_\perp \xi_5^{-1})^{\eta_C} & \ \mbox{if} \ &  \ (\brm{q}_\perp,
q_d) = (\brm{0}, 0)
\end{array}
\right. \nonumber
\end{eqnarray}
and
\begin{eqnarray}
\label{revCSEK}
K \sim  \left\{
\begin{array}{lcl}
(L_\perp |\brm{q}_\perp|)^{-\eta_K} &\ \mbox{if} \ & \ \xi_5
|\brm{q}_\perp| \gg 1\, , \ q_d =0
\\
(L_d q_d)^{-\eta_K/\phi} &\ \mbox{if} \ & \ \xi_5  |\brm{q}_\perp|
L_d /L_\perp\gg 1 \, , \ \brm{q}_\perp = \brm{0}
\\
(L_\perp \xi_5^{-1})^{-\eta_K} & \ \mbox{if} \ &  \
(\brm{q}_\perp, q_d) = (\brm{0}, 0)
\end{array}
\right.
\nonumber\\
\end{eqnarray}
where $\xi_5$ is a correlation length given by
\begin{eqnarray}
\xi_5 = L_\perp\,  ( L_\perp^2 C_5/K)^{-\nu_5} \, .
\end{eqnarray}
The elastic constant $C_1$ is not renormalized and it is not
singular in either the wavenumbers or $\xi_5$.  When $C_5$ is
nonzero, $\Gamma_{dd} \sim C_5^{\gamma_5}q_{\perp}^2 + K
q_{\perp}^4$ at small $\brm{q}$, where $\gamma_5 = (2 - \eta_K)
\nu_5$ and $K$ is given by the last expression in Eq.\
(\ref{revCSEK}).

At exactly three dimensions the above power law singularities
become logarithmic singularities
\begin{subequations}
\begin{eqnarray}
C_2 &\sim& C_3 \sim C_4 \sim | \ln (|\brm{q}_\perp|/\mu)|^{-1/7}
\, ,
\\
K &\sim& |\ln  (|\brm{q}_\perp|/\mu)|^{4/7} \, ,
\end{eqnarray}
\end{subequations}
where $\mu$ is a wavenumber scale. This logarithmic anomaly can be
observed for $\xi_\perp |\brm{q}_\perp| \ll 1$, where
\begin{eqnarray}
\xi_\perp = \mu^{-1} \exp \left[32 \pi \sqrt{C_1 K^3}/(7 T
C_4)\right] \, .
\end{eqnarray}

In the critical regime at small $C_5$, we find the following
universal Poisson ratios:
\begin{eqnarray}
C_2^2/(C_1 C_4) = 0 \quad \mbox{and} \quad C_3/C_4 =  -1/2 \, .
\end{eqnarray}

\subsection{Nematic elastomers}
At the transition from the isotropic to the nematic phase,
liquid-crystalline elastomers undergo an anisotropic stretch
relative to their isotropic reference state of a factor
$\Lambda_{0||}$ along the nematic axis and $\Lambda_{0\perp}$
perpendicular to it.  In the absence of explicit uniaxial terms
such as
\begin{eqnarray}
\label{breakingTerm}
h \int d^{d_\perp} x_\perp \int dx_d \left[ u_{dd} - \frac{1}{d}
u_{ii} \right] \, ,
\end{eqnarray}
the elastic energy of the nematic phase that forms spontaneously
from an isotropic phase is soft, i.e., the elastic constant $C_5$
for shears in the uniaxial plane vanishes.  After rescaling
lengths measured relative to the anisotropic nematic reference
state and displacements according to $x_a \rightarrow x_a$, $x_d
\rightarrow \sqrt{r-1} \, x_d$, $u_a \rightarrow u_a$ and $u_d
\rightarrow \sqrt{r-1} \, u_d$, where $r=
\Lambda_{0||}^2/\Lambda_{0\perp}^2$, the relevant parts of the
elastic energy can be written when uniaxial terms are present and
small as
\begin{eqnarray}
\mathcal{H} &=& \frac{1}{2} \int d^{d_\perp} x_\perp \int d x_d \,
\big\{  C_1\,  v_{dd}^2 + K \left(  \nabla_\perp^2 u_d \right)^2
 \\
&+& 2 \, C_2 \, v_{dd} v_{aa} +  C_3 \, v_{aa}^2 + 2 \, C_4 \,
v_{ab}^2 +  C_5 \, v_{ad}^2 \big\}  \, , \nonumber
\end{eqnarray}
with nonstandard strains $v_{ab} = \frac{1}{2} (  \partial_a  u_b
+  \partial_b  u_a - \partial_a  u_d \partial_b  u_d )$, $v_{dd} =
\partial_d  u_d + \frac{1}{2}  \partial_a  u_d \partial_b  u_d$
and $v_{ad} =  \frac{1}{2} \partial_a  u_d$ and where $C_5$ goes
linearly to zero with $h$ or the magnitude of other uniaxial
terms.

The scaling of the inverse displacement correlation functions of
NEs is similar to but not identical to that of the CSE model:
\begin{subequations}
\label{NEscaleForms}
\begin{eqnarray}
\Gamma_{dd}(\brm{q}_{\perp} , q_d) &= &\frac{K}{T} L_{\perp}^{-4}
\ell^{4- \eta_K}
 \\
&\times&  {\hat \Phi}_{dd} \left(
\frac{L_{\perp}\brm{q}_{\perp}}{\ell^{1}} , \frac{L_d
q_d}{\ell^{\phi}}, \frac{C_5}{K} \frac{L_{\perp}^2}{\ell^{1/\nu_h
}}, \frac{C_1}{C_4} \frac{1}{\ell^{\eta_C}}  \right) \, ,
\nonumber \\
\Gamma_{ad}(\brm{q}_{\perp} , q_d) &= &\frac{C_2}{T}
L_{\perp}^{-1} L_d^{-1} \ell^{1+ \phi}
 \\
&\times&  {\hat \Phi}_{ad} \left(
\frac{L_{\perp}\brm{q}_{\perp}}{\ell^{1}} , \frac{L_d
q_d}{\ell^{\phi}}, \frac{C_5}{K} \frac{L_{\perp}^2}{\ell^{1/\nu_h
}}  \right) \, ,
\nonumber \\
\Gamma_{ab}(\brm{q}_{\perp} , q_d)&=&
\frac{C_3}{T}L_{\perp}^{-2}\ell^{2}
\\
&\times& {\hat \Phi}_{ab}
 \left( \frac{L_{\perp}\brm{q}_{\perp}}{\ell^{1}} , \frac{L_d q_d}{\ell^{\phi}}, \frac{C_5}{K} \frac{L_{\perp}^2}{\ell^{1/\nu_h }}, \frac{C_4}{C_3} \, \ell^{\eta_C}  \right) \, ,
\nonumber
\end{eqnarray}
\end{subequations}
where, to first order in $\varepsilon$
\begin{subequations}
\begin{eqnarray}
&&\eta_K = 38 \epsilon/59 \, ,\quad \eta_C = 4\epsilon/59 \,  ,
\\
&&\phi = 2 - 21\varepsilon /59 \, ,\quad \nu_h = 1/2 +
9\varepsilon/108 \,  ,
\end{eqnarray}
\end{subequations}
and
\begin{subequations}
\begin{eqnarray}
L_\perp &\sim& \left[ \sqrt{K^3 /(C_4 \, T^2)}
\right]^{1/\varepsilon}  ,
\\
L_d &=& \sqrt{C_4/K} \, L_\perp^2 \, .
\end{eqnarray}
\end{subequations}
Note that four independent scaling exponents, $\eta_K $, $\eta_C$,
$\phi$, and $\nu_h$ are required to describe NEs with a small
uniaxial energy.  In the above, lengths, displacements and
$\brm{q}$-vectors are measured in rescaled units.

The above scaling forms predict that $C_1$, $C_2$, and $C_3$ are
unrenormalized and that
\begin{eqnarray}
\label{finResC4NE}
C_4 \sim  \left\{
\begin{array}{lcl}
(L_\perp |\brm{q}_\perp|)^{\eta_C} &\ \mbox{if} \ & \ \xi_h
|\brm{q}_\perp| \gg 1\, , \ q_d =0
\\
(L_d q_d)^{\eta_C/\phi} &\ \mbox{if} \ & \ \xi_h  |\brm{q}_\perp|
L_d /L_\perp\gg 1 \, , \ \brm{q}_\perp = \brm{0}
\\
(L_\perp \xi_h^{-1})^{\eta_C} & \ \mbox{if} \ &  \ (\brm{q}_\perp,
q_d) = (\brm{0}, 0)
\end{array}
\right.
\nonumber\\
\end{eqnarray}
as well as
\begin{eqnarray}
\label{finResKNE}
K \sim  \left\{
\begin{array}{lcl}
(L_\perp |\brm{q}_\perp|)^{-\eta_K} &\ \mbox{if} \ & \ \xi_h
|\brm{q}_\perp| \gg 1\, , \ q_d =0
\\
(L_d q_d)^{-\eta_K/\phi} &\ \mbox{if} \ & \ \xi_h  |\brm{q}_\perp|
L_d /L_\perp\gg 1 \, , \ \brm{q}_\perp = \brm{0}
\\
(L_\perp \xi_h^{-1})^{-\eta_K} & \ \mbox{if} \ &  \
(\brm{q}_\perp, q_d) = (\brm{0}, 0)
\end{array}
\right.
\nonumber\\
\end{eqnarray}
with $\xi_h$ given by
\begin{eqnarray}
\xi_h = L_\perp\,  ( L_\perp^2 h/K)^{-\nu_h} \, .
\end{eqnarray}
At small but nonzero $h$, $\Gamma_{dd} \sim h^{\gamma_h}
q_{\perp}^2 + K q_{\perp}^4$ at small $\brm{q}$, where $\gamma_h =
( 2 - \eta_h) \nu_h$ and $K$ is given by the last expression in
Eq.\ (\ref{finResKNE}).

At exactly three dimensions the above power laws become
\begin{subequations}
\begin{eqnarray}
C_4 &\sim& |\ln(|\brm{q}_\perp|/\mu)|^{-4/59} \, ,
\\
K &\sim& |\ln (|\brm{q}_\perp|/\mu)|^{38/59} \, .
\end{eqnarray}
\end{subequations}
The length scale that marks the crossover from harmonic to
logarithmic behavior is
\begin{eqnarray}
\xi_\perp = \mu^{-1} \exp \left[64 \pi \sqrt{ K^3}/(7 \sqrt{6 C_4}
T ) \right] \, .
\end{eqnarray}

Provided that $C_5$ is small, the critical regime entails four
independent Poisson ratios:
\begin{subequations}
\begin{eqnarray}
&&C_2/C_1 =1\, , \  \ C_3/C_1 =1\, , \ \ C_4/C_1 =0\, ,
\\
&& (2C_2 - C_3 -C_1)/C_4 = 1/2 \, .
\end{eqnarray}
\end{subequations}

\section{Critically soft elastomers}
\label{ise}

\subsection{The model}
\label{iseModel}
We start by setting up a field theoretic minimal model for CSEs
that is suitable for our subsequent RG analysis. We find it
convenient to use the Lagrangian formulation of
elasticity~\cite{landau_lifshitz_elasticity,tomsBook}. In this
formulation the mass points of the equilibrium undistorted medium
are labeled by their position vectors $\brm{x}$ in $d$-dimensional
(reference) space. When the medium is distorted, a mass point
originally at $\brm{x}$ is mapped to a new point $\brm{R}
(\brm{x})$ in $d$-dimensional (target) space. Since $\brm{R}
(\brm{x}) = \brm{x}$ when there is no distortion, it is customary
to introduce the phonon variable $\brm{u} (\brm{x}) = \brm{R}
(\brm{x}) - \brm{x}$ that measures the deviation of $\brm{R}
(\brm{x})$ from $\brm{x}$.

Suppose for a moment the medium is distorted solely by stretching.
The energy of the distorted state relative to the reference state
depends on the relative amount of stretching $d \brm{R}^2 - d
\brm{x}^2 = 2 u_{ij} d x_i d x_j $, where
\begin{eqnarray}
u_{ij} = \frac{1}{2} \left\{ \partial_i  u_j + \partial_j u_i +
\partial_i  u_k  \partial_j  u_k  \right\}
\end{eqnarray}
with $i, j, k = 1, \ldots , d$~\cite{footnote1} are the components
of the familiar nonlinear Lagrangian strain tensor $\tens{u}$.
Note that $\tens{u}$ is invariant under arbitrary rotations in
target space. This feature makes the Lagrangian strain tensor an
adequate variable for formulating elastic energies because all
elastic media are rotationally invariant in target space. This
invariance is easy to understand: different physical orientations
of the same sample have the same energy~\cite{footnote2}. As
customary, the reference state, relative to which $\tens{u}$ is
defined, is taken to be in mechanical equilibrium, and hence no
terms linear in $u_{ij}$ appear in the stretching energy. To
lowest order, the stretching energy is then of the form
\begin{eqnarray}
H_{\text{st}} = \frac{1}{2} \int d^d x K_{ijkl} u_{ij} u_{kl} \, ,
\end{eqnarray}
where $K_{ijkl}$ is an elastic constant tensor. For media
isotropic in the reference space, for example, there are only two
independent elastic constants in $K_{ijkl}$ that are known as the
Lam\'{e} coefficients $\lambda$ and $\mu$. Media with uniaxial
symmetry in the reference space are characterized in general by
five independent elastic constants. Assuming that the anisotropy
axis is in the $\hat{\brm{e}}_d = (0, \ldots ,1)$ direction we may
write the stretching energy as
\begin{eqnarray}
\label{elasten}
&&H_{\text{st}} =  \frac{1}{2} \int d^{d_\perp} x_\perp \int d x_d
\big\{ C_1 u_{dd}^2 + 2 \, C_2 u_{dd} u_{aa} \nonumber \\
&&+ \,  C_3  u_{aa}^2 + 2  C_4 u_{ab}^2 + C_5 u_{ad}^2 \big\} \
\end{eqnarray}
with $d_\perp = d-1$ and $a,b = 1, \ldots , d_\perp$.

Now suppose that the elastic constant $C_5$ vanishes.  Rewriting
$H_{\text{st}}$ in Fourier space one sees easily that the
stretching energy cost is zero for phonon displacements
$\widetilde{\brm{u}} (\brm{q})$ perpendicular to $\hat{\brm{e}}_d$
with momentum $\brm{q}$ parallel to $\hat{\brm{e}}_d$ and for
$\widetilde{\brm{u}} (\brm{q})$ parallel to $\hat{\brm{e}}_d$ with
$\brm{q}$ perpendicular to $\hat{\brm{e}}_d$. In other words: CSEs
are soft elastic materials.

For many elastic systems it is justified to neglect energetic
contributions, such as bending, that are associated with higher
derivatives of the displacements. That is, because bending is
unimportant compared to stretching at small momenta. Due to the
soft elasticity, however, the stretching energy of CSEs can
vanish, and hence, bending is important.

For the moment, we set aside the uniaxial term proportional to
$C_5$ and concentrate on the pure soft case. The effects of a
small but non-vanishing $C_5$ will be included later on. Taking
into account stretching and bending, the CSE model is defined by
the Hamiltonian
\begin{eqnarray}
\label{Hamil}
&&\mathcal{H} = \frac{1}{2} \int d^{d_\perp} x_\perp \int d x_d \,
\Big\{  C_1 \, u_{dd}^2 + K \left(  \nabla_\perp^2 u_d \right)^2
\nonumber \\
&&   + \, 2 C_2 \,  u_{dd} u_{aa}    +  C_3 \, u_{aa}^2 + 2C_4 \,
u_{ab} u_{ab} \Big\} \, ,
\end{eqnarray}
where $K$ is a bending modulus. All other bending terms allowed by
symmetry turn out to be irrelevant in the sense of the
renormalization group. Also, not all parts of the strains are
relevant. Discarding  any parts of the strains that lead to
irrelevant contributions to the Hamiltonian, as discussed further
below, leaves us with
\begin{subequations}
\label{truncStrains}
\begin{eqnarray}
u_{ab} = \frac{1}{2} \left\{ \partial_a u_b + \partial_b u_a +
\partial_a u_d \partial_b u_d \right\}
\end{eqnarray}
and
\begin{eqnarray}
u_{dd} = \partial_d u_d \, .
\end{eqnarray}
\end{subequations}
In principle we could use $\mathcal{H}$ as it stands in
Eq.~(\ref{Hamil}) for our RG analysis. We find it convenient,
however, to reduce the number of constants featured in
$\mathcal{H}$ at the onset. To this end, we rescale $u_a \to
(K/C_4) u_a$,  $u_d \to \sqrt{K/C_4} u_d$, and $x_d \to (C_4/K^2)
x_d$. Then the Hamiltonian takes on the form
\begin{eqnarray}
\label{fieldHamil}
&&\mathcal{H} = \frac{1}{2} \int d^{d_\perp} x_\perp \int d x_d \,
\Big\{  \omega \, u_{dd}^2 +  \left(  \nabla_\perp^2 u_d \right)^2
\nonumber \\
&&   + \, 2 g \,  u_{dd} u_{aa}    +  f \, u_{aa}^2 + 2 \, u_{ab}
u_{ab} \Big\} \, ,
\end{eqnarray}
where
\begin{equation}
\omega = C_1 K^3 / C_4^2 \, ,\quad g = C_2 (K / C_4)^{3/2} \, ,
\quad f = C_3 / C_4 .
\end{equation}
At this stage we would like to point out that we explicitly keep
the temperature $T$ in the Boltzmann weight $\exp (- \mathcal{H} /
T)$~\cite{footnoteKb} governing our field theoretic calculations.
In what follows we carry out a perturbation expansion in the
temperature, i.e., $T$ serves as our expansion parameter. As a
consequence, not only the constants and fields featured in
$\mathcal{H}$ but also the temperature will require
renormalization.

An effective Hamiltonian for $u_d$ alone can be obtained by
integrating out the transverse variable $u_a$ from the full CSE
Hamiltonian of Eq.\ (\ref{Hamil}).  When $C_4=\infty$, this
process leads to the Hamiltonian for a smectic-$A$ liquid crystal
whose anomalous elasticity was analyzed by Grinstein and
Pelcovits~\cite{grinstein_pelcovits_81_82}.  Our rescaling of
variables to obtain Eq.\ (\ref{fieldHamil}) with the coefficients
of both $(\nabla_{\perp}^2 u_d )^2$ and $2 u_{ab} u_{ab}$ set to
unity is not ideally suited to taking the $C_4 = \infty$ limit.
Our primary interest is the anomalous elasticity unique to soft
uniaxial systems for which the parametrization of Eq.\
(\ref{fieldHamil}) is appropriate. We will not give further
consideration to the Grinstein-Pelcovits limit of our model.

As a further step towards our RG analysis, we now discuss the
scaling symmetries of our model. First, under a global rescaling
of the coordinates $x_a \to \mu^{-1} x_a$ and  $x_d \to \mu^{-2}
x_d$, we find a scaling invariant theory provided that $u_a \to
\mu^{1} u_a$ and $T \to \mu^{3-d} T$ ($\mu$ invariance). Viewing
$\mu$ as an inverse length scale, this means that the field $u_a$
has a naive dimension 1 and that the naive dimension of $T$ is
$\varepsilon = 3-d$. The field $u_d$ and the remaining parameters
in $\mathcal{H}$ have a vanishing naive dimension. Above $d=3$
dimensions the naive dimension of $T$ is negative and $T$ is
irrelevant whereas it is relevant below $d=3$. Hence, we identify
$d_c =3$ as the upper critical dimension of the CSE model.

At this point we take a short detour and catch up on justifying
the truncation of the strains as stated in
Eqs.~(\ref{truncStrains}). Applying the $\mu$-rescaling to the
original full strains leads to
\begin{subequations}
\label{truncStrainsJust}
\begin{eqnarray}
u_{ab} \to \frac{\mu^2}{2} \left\{ \partial_a u_b + \partial_b u_a
+ \partial_a u_d \partial_b u_d + \mu^2 \partial_a u_c \partial_b
u_c \right\}
\nonumber \\
\end{eqnarray}
and
\begin{eqnarray}
u_{dd} \to \mu^2 \left\{  \partial_d u_d  + \frac{\mu^2}{2}
(\partial_d u_d)^2 + \frac{\mu^4}{2} \partial_d u_c \partial_d u_c
\right\} .
\end{eqnarray}
\end{subequations}
The terms carrying extra powers of $\mu$ do not contribute to the
leading behavior in the limit $\mu \to 0$. Hence, they can be
neglected in studying the long length scale behavior at leading
order, i.e., they are irrelevant in the sense of the RG.

Second, due to the anisotropy of the model, we may rescale the
longitudinal coordinate alone: $x_d \to \beta x_d$. Scale
invariance is retained if $\omega \to \beta^2 \omega$, $g \to
\beta g$, and $T \to \beta T$ ($\beta$ invariance). Note that the
composed couplings
\begin{eqnarray}
\sigma = g^2 / \omega , \ \rho = f , \
 \mbox{and} \ \ t = \mu^{-\varepsilon} T/\sqrt{\omega}
\end{eqnarray}
are invariant under the longitudinal rescaling. The factor
$\mu^{-\varepsilon}$ is included in the definition of $t$ to
render it, like $\sigma$ and $\rho$, dimensionless. As we go
along, we will see that $\sigma$, $\rho$, and $t$ emerge quite
naturally in perturbation theory. Third, $\mathcal{H}$ is
invariant under the rescaling $u_d \to u_d + f_d (\brm{x}_\perp
)$, where $f_d$ is an arbitrary function of the transversal
coordinate $\brm{x}_\perp$. Fourth, rescaling $u_a \to u_a + f_a
(x_d ) + M_{ab} x_b$ leaves $\mathcal{H}$ invariant if the matrix
constituted by the $ M_{ab}$ is antisymmetric and $f_a$ is a
function of the longitudinal coordinate only. Finally,
$\mathcal{H}$ is invariant under the transformation $u_a \to u_a +
\theta_d u_d$ and  $u_d \to u_d - \theta_a x_a$ provided that the
$\theta$'s are small. Note that this transformation mixes the
longitudinal and the transversal fields (mixing invariance). It
can be viewed as a remnant of the rotational invariance of the
original theory in target space. This mixing transformation will
be valuable for us because it leads to Ward identities that reduce
the number of vertex functions to be calculated in perturbation
theory. These Ward identities will be derived in
Appendix~\ref{app:ward1}.

\subsection{Renormalization group analysis}
\label{iserg}
In this section we determine the scaling behavior of the
correlation function of the fields $u_a (\brm{x})$ and $u_d
(\brm{x})$ by using perturbation theory augmented by
renormalization group methods. As usual, we analyze vertex
functions that require renormalization due to the presence of
ultraviolet (UV) divergences in Feynman diagrams. Our main tools
in this section will be dimensional regularization and minimal
subtraction. To avoid infrared (IR) singularities in the Feynman
diagrams, we supplement our Hamiltonian with a mass term,
\begin{eqnarray}
\mathcal{H} \to \mathcal{H} + \frac{\tau}{2} \int d^{d_\perp}
x_\perp \int d x_d  \,  u_{d}^2 \, .
\end{eqnarray}
At the appropriate stage of the calculations we then sent $\tau$
to zero to recover the original situation.

\subsubsection{Diagrammatic expansion}
In order to set up a diagrammatic perturbation expansion we have
to determine its constituting elements. First, we have the
Gaussian propagator $\underline{\underline{G}}$ that has the form
of a $d \times d$ matrix. The elements of its inverse
$\underline{\underline{\Gamma}}$ are readily collected from the
Hamiltonian:
\begin{subequations}
\label{GaussVertices}
\begin{eqnarray}
\Gamma_{dd} &=& T^{-1} \big[ \tau +\omega q_d^2 +  \brm{q}_\perp^4
\big] \, ,
\\
\Gamma_{ad} &=& T^{-1} g q_a q_d \, ,
\\
\Gamma_{ab} &=& T^{-1} \big[ (f+1) q_a q_b + \delta_{ab}
\brm{q}_\perp^2 \big] \,  .
\end{eqnarray}
\end{subequations}
Inverting $\underline{\underline{\Gamma}}$ we find that the
Gaussian propagator has the elements
\begin{subequations}
\label{GaussProp}
\begin{eqnarray}
G_{dd} &=&T \,  \frac{B}{B \tau +A q_d^2 + B \brm{q}_\perp^4}  \,
,
\\
G_{ad} &=& T\,  \frac{-g}{B \tau +A q_d^2 + B \brm{q}_\perp^4} \,
\frac{q_a q_d}{\brm{q}_\perp^2} \, ,
\\
G_{ab} &=&  T \left[ \frac{\delta_{ab}}{\brm{q}_\perp^2}  -
\frac{D \tau +C q_d^2 + D \brm{q}_\perp^4}{B \tau +A q_d^2 + B
\brm{q}_\perp^4} \, \frac{q_a q_b}{\brm{q}_\perp^4} \right]  ,
\end{eqnarray}
\end{subequations}
where we have used the shorthand notations $A = \omega (f+2) -
g^2$, $B = f+2$, $C = \omega (f+1) - g^2$, and $D = f+1$. Second,
our diagrammatic expansion features the 4 vertices
\begin{subequations}
\label{CSEvertices}
\begin{eqnarray}
&&i\,  \frac{g}{2 \, T} \, q_d^{(1)}  q_b^{(2)} q_b^{(3)}  \, ,
\\
&&i \, \frac{f}{2 \, T} \, q_a^{(1)}  q_b^{(2)} q_b^{(3)}  \, ,
\\
&&i \,  \frac{1}{T} \, q_a^{(2)}  q_b^{(1)} q_b^{(3)}   \, ,
\\
&&- \, \frac{f+2}{8 \, T} \, q_a^{(1)}  q_a^{(2)} q_b^{(3)}
q_b^{(4)}   \, .
\end{eqnarray}
\end{subequations}
It is understood that the sum of the momenta has to vanish at each
vertex.

Next, we need to determine which of the vertex functions
$\Gamma^{(M,N)}$ with $M$ external $u_a$-legs and $N$ external
$u_d$-legs are superficially UV divergent. Analyzing their
topology, we find that the superficial degree of divergence
$\delta$ of our diagrams is given at the upper critical dimension
by $\delta = 4 - M - 2D_\parallel - D_\perp$, where $D_\parallel$
($D_\perp$) is the number of longitudinal (transversal )
derivatives on the external legs. Thus, the only vertex functions
containing superficially divergent diagrams are
$\Gamma^{(0,1)}_{d}$, $\Gamma^{(1,0)}_{a}$, $\Gamma^{(0,2)}_{dd}$,
$\Gamma^{(1,1)}_{ad}$, $\Gamma^{(2,0)}_{ab}$,
$\Gamma^{(0,3)}_{ddd}$, $\Gamma^{(1,2)}_{add}$, and
$\Gamma^{(0,4)}_{dddd}$. All these vertex functions have to be
taken into account in the renormalization procedure. By virtue of
the mixing invariance, however, there exist several relations
between the vertex functions in form of Ward identities. These are
derived and stated in Appendix~\ref{app:ward1}. Due to these Ward
identities it is sufficient for our purposes to actually calculate
the two-point functions $\Gamma^{(0,2)}_{dd}$,
$\Gamma^{(1,1)}_{ad}$, and $\Gamma^{(2,0)}_{ab}$. Once the
equations of state $\Gamma^{(0,1)}_{d} = 0$ and
$\Gamma^{(1,0)}_{a} = 0$ are satisfied and the 2-point functions
are renormalized, the Ward identities guarantee that the remaining
vertex functions are cured of their UV divergences.

We calculate the two-point vertex functions to one-loop order
using dimensional regularization. The Feynman diagrams entering
this calculation are listed in Figs.~\ref{diagrams1} to
\ref{diagrams3}. Details on computing the diagrams can be found in
Appendix~\ref{app:diagramCalc}.
\begin{figure}
\includegraphics[width=8.6cm]{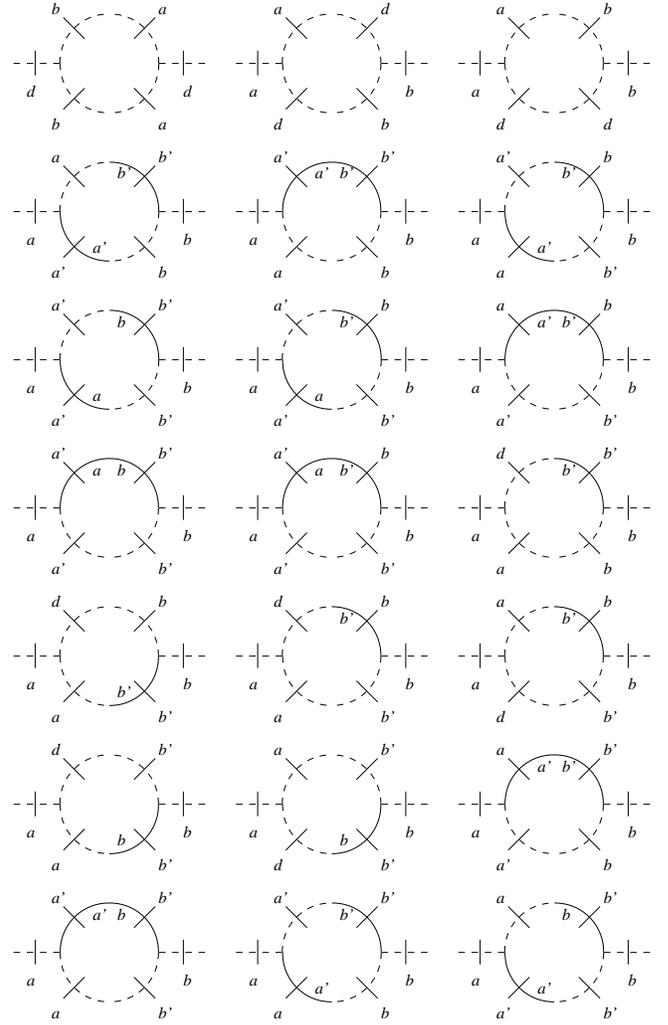}
\caption[]{\label{diagrams1}Feynman diagrams contributing to $\Gamma_{dd}$. The dashed lines symbolize the element $G_{dd}$ of the Gaussian propagator. Lines half dashed and half solid with an index, say $a$, stand for $G_{ad}$. Solid lines with 2 indices, say $a$ and $b$, visualize $G_{ab}$. The ticks indicate derivatives with respect to the reference space coordinate with an index specifying the component.}
\end{figure}
\begin{figure}
\includegraphics[width=5.6cm]{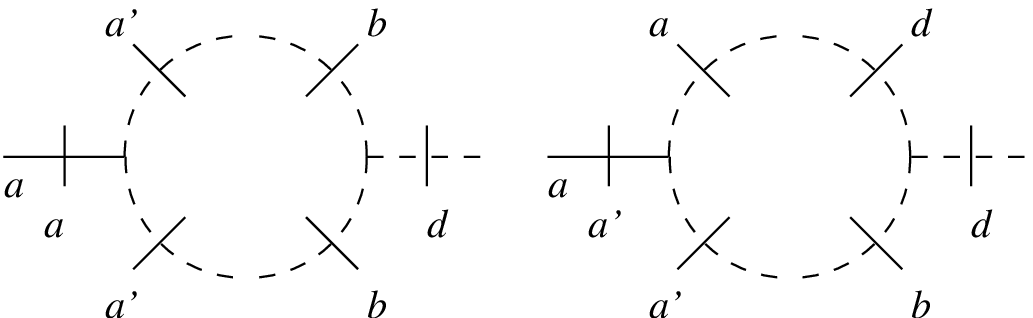}
\caption[]{\label{diagrams2}Feynman diagrams contributing to $\Gamma_{ad}$. The meaning of the symbols is the same as in Fig.~\ref{diagrams1}.}
\end{figure}
\begin{figure}
\includegraphics[width=8.6cm]{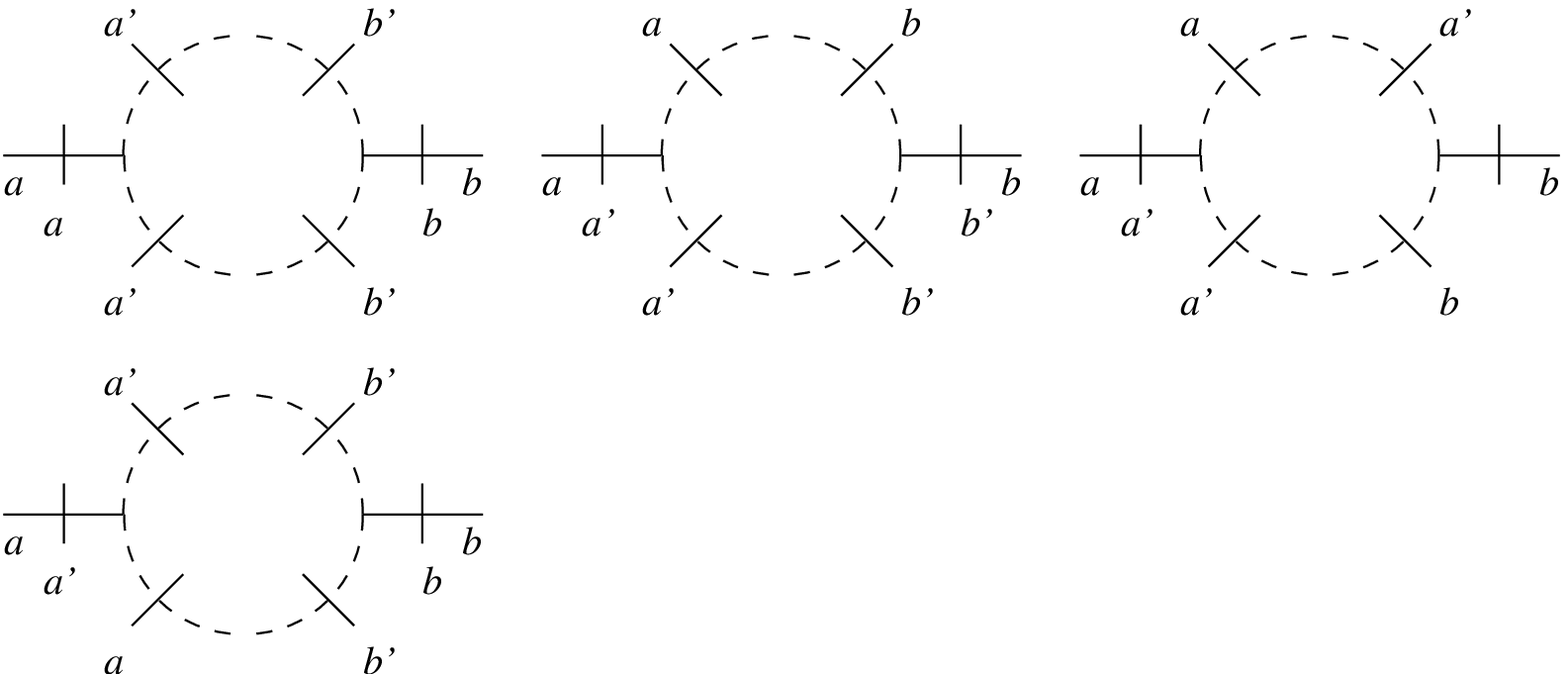}
\caption[]{\label{diagrams3}Feynman diagrams contributing to $\Gamma_{ab}$. The meaning of the symbols is the same as in Fig.~\ref{diagrams1}.}
\end{figure}
Our results for the 2-point functions read
\begin{subequations}
\label{vertexFkts}
\begin{eqnarray}
&&\Gamma^{(0,2)}_{dd} = T^{-1} \big[ \tau +\omega q_d^2 +
\brm{q}_\perp^4 \big]
 \\
& &-\,  \frac{{\sqrt{2 + f}}\,g^2\, q_d^2 }
  {16\,\pi \,\varepsilon \,{\sqrt{-g^2 + \left( 2 + f \right) \omega }}} \, \tau^{-\varepsilon /4 }\, ,
  \nonumber
  \\
&& -\, \frac{3\, (1+f)\, \brm{q}_\perp^4}{8\,\pi \,\varepsilon \,
    {\sqrt{2 + f}} \, \sqrt{-g^2 + \left( 2 + f \right) \omega}} \, \tau^{-\varepsilon /4 }\, ,
\nonumber \\
&&\Gamma^{(1,1)}_{ad} = T^{-1} g \, q_a q_d
\\
&& -\, \frac{ {\sqrt{2 + f}}\, \left( 1 + f \right)\,g\,q_d\,
      {q_a} }{16\,\pi \,\varepsilon \,
    {\sqrt{-g^2 + \left( 2 + f \right) \omega }}} \, \tau^{-\varepsilon /4 } \, ,
\nonumber
\\
&&\Gamma^{(2,0)}_{ab} = T^{-1} \big[ (f+1) \, q_a q_b +
\delta_{ab} \, \brm{q}_\perp^2 \big]
 \\
&&-\, \frac{ {\sqrt{2 + f}}\,[ 2\,{\left( 1 + f \right) }^2\,{q_a}
         {q_b} + {{\delta }_{ab}} \, \brm{q}_\perp^2]   }{32\,\pi \,
    \varepsilon \,{\sqrt{-g^2 + \left( 2 + f \right) \omega }}} \,  \tau^{-\varepsilon /4 } \, .
    \nonumber
\end{eqnarray}
\end{subequations}
As already indicated by the vertex functions' superficial degree
of divergence, higher order terms in the momentum expansion are
convergent and hence can be neglected for our purposes. Likewise,
contributions proportional to the mass $\tau$ are convergent.
Thus, $\tau$ does not require renormalization and
consequently its scaling dimension is identical to its naive
dimension four. At this stage of the calculation $\tau$ has
fulfilled its purpose, viz.\ it prevented IR singularities from
producing spurious $\varepsilon$ poles. In conjunction with the
$\varepsilon$ expansion we now can safely send $\tau$ to zero.

\subsubsection{Renormalization}
The UV divergences have their manifestation in the $\varepsilon$
poles appearing in Eqs.~(\ref{vertexFkts}). We eliminate these
poles by employing the renormalization scheme
\begin{subequations}
\begin{eqnarray}
&&u_d \to \mathaccent"7017{u}_d  = Z^{1/2} u_d \, ,
\\
&&u_a \to \mathaccent"7017{u}_a  = Z u_a \, ,
\\
&&T \to \mathaccent"7017{T} = Z Z^{-1}_T T  \, ,
\\
&&\omega \to \mathaccent"7017{\omega} = Z_T^{-1} Z_\omega \omega
\, ,
\\
&&g \to \mathaccent"7017{g} = Z^{-1/2}  Z_T^{-1} Z_g g \, ,
\\
&&f \to \mathaccent"7017{f} = Z^{-1}  Z_T^{-1} Z_f f \, ,
\end{eqnarray}
\end{subequations}
where the $\mathaccent"7017{}$ indicates unrenormalized
quantities. Our scheme is chosen so that the Hamiltonian retains
its original structure:
\begin{eqnarray}
\label{renoHamil}
&&\frac{\mathcal{H}}{T}  \to \frac{1}{2 \, T} \int d^{d_\perp}
x_\perp \int d x_d  \, \Big\{   Z_\omega \omega \, u_{dd}^2 +  Z_T
\left(  \nabla_\perp^2 u_d \right)^2
\nonumber \\
&&   + \, 2 Z_g g \,  u_{dd} u_{aa}    +  Z_f f \, u_{aa}^2 + 2 \,
Z_T Z u_{ab} u_{ab} \Big\} \, ,
\end{eqnarray}
The simplest way of determining the renormalization Z-factors is
minimal subtraction. In this procedure the Z-factors are chosen so
that they solely cancel the $\varepsilon$ poles and otherwise
leave the vertex functions unchanged. Expressed in terms of the
effective couplings introduced in Sec.~\ref{iseModel} our
Z-factors are of the structure
\begin{equation}
\label{expZ}
Z_{\ldots }(t, \sigma ,\rho ) =1+\sum_{m=1}^{\infty
}\frac{X_{\ldots }^{(m)}(t, \sigma ,\rho )}{\varepsilon^{m}}\, .
\end{equation}
The $X_{\ldots }^{(m)}(t, \sigma ,\rho )$ are expansions in the
effective temperature $t$ beginning with the power $t^{m}$. It is
a fundamental fact of renormalization theory,
cf.~Ref.~\cite{amit_zinn-justin}, that this procedure is suitable
to eliminate all the UV-divergences (not only the superficial
ones) from any vertex function order by order in perturbation
theory. To one-loop order we find that our Z factors are given by
\begin{subequations}
\label{CSEZs}
\begin{eqnarray}
&&Z  = 1 +  t \, \frac{14 + 13 \rho}{ 32 \, \pi \, \varepsilon \,
\sqrt{2+\rho} \, \sqrt{2+\rho - \sigma} } \, ,
\\
&&Z_T =  1 -  t \, \frac{3 \, (1+\rho)}{8\, \pi \, \varepsilon \,
\sqrt{2+\rho} \, \sqrt{2+\rho - \sigma} } \, ,
\\
&&Z_\omega  = 1 + t \, \frac{\sqrt{2+\rho} \, \sigma}{16 \, \pi \,
\varepsilon \, \sqrt{2+\rho - \sigma}} \, ,
\\
&&Z_g = 1 + t \, \frac{\sqrt{2+\rho} \, (1+\rho)}{8 \, \pi \,
\varepsilon \, \sqrt{2+\rho - \sigma}}  \, ,
\\
&&Z_f = 1 + t \, \frac{\sqrt{2+\rho} \, (1+4\rho + 2\rho^2)}{32 \,
\pi \, \varepsilon \, \sqrt{2+\rho - \sigma}} \, .
\end{eqnarray}
\end{subequations}

\subsubsection{Scaling 1: RG equation and its solution}
\label{CSERG}
Next, we infer the scaling behavior of vertex function from a RG
equation. This RG equation is a manifestation of the fact that the
unrenormalized theory has to be independent of the arbitrary
length scale $\mu^{-1}$ introduced by renormalization. By virtue
of this independence, the unrenormalized vertex functions satisfy
the identity
\begin{eqnarray}
\label{independence}
\mu \frac{\partial}{\partial \mu} \mathaccent"7017{\Gamma}^{(M,N)}
\left( \left\{ {\rm{\bf q}}_\perp ,q_d \right\} ;
\mathaccent"7017{\omega}, \mathaccent"7017{T},
\mathaccent"7017{g},  \mathaccent"7017{f} \right) = 0 \, .
\end{eqnarray} The identity~(\ref{independence}) translates via
the Wilson functions~\cite{footnoteWilson}
\begin{subequations}
\label{wilson}
\begin{eqnarray}
\label{wilsonGamma}
&&\gamma_{\ldots }  = \mu \frac{\partial }{\partial \mu} \ln
Z_{\ldots } \bigg|_0 \, ,
\\
\label{wilsonZeta}
&&\zeta  = \mu \frac{\partial \ln \omega}{\partial \mu} \bigg|_0 =
\gamma_T - \gamma_\omega \, ,
\\
&&\beta_t = \mu \frac{\partial t}{\partial \mu} \bigg|_0 = t \,
\bigg( - \varepsilon - \gamma + \frac{1}{2} \gamma_T + \frac{1}{2}
\gamma_\omega  \bigg) \, , \quad
\\
\label{betasigma}
&&\beta_\sigma = \mu \frac{\partial \sigma}{\partial \mu} \bigg|_0
= \sigma \, \left(  \gamma +  \gamma_T +  \gamma_\omega  - 2
\gamma_g \right) \, ,
\\
\label{betarho}
&&\beta_\rho = \mu \frac{\partial \rho}{\partial \mu} \bigg|_0 =
\rho \, \left(  \gamma +  \gamma_T -  \gamma_f   \right)
\end{eqnarray}
\end{subequations}
into the Gell-Mann-Low RG equation
\begin{eqnarray}
\left[ D_\mu -  \left( M + \frac{N}{2} \right) \gamma \right]
\Gamma^{(M,N)} \left( \left\{ {\rm{\bf q}}_\perp ,q_d \right\} ;
\omega , t, \sigma , \rho , \mu \right) = 0 \, .
\nonumber \\
\end{eqnarray}
Here we have used the shorthand notation
\begin{eqnarray}
\label{rgOp}
D_\mu = \mu \frac{\partial }{\partial \mu} + \omega \zeta
\frac{\partial }{\partial \omega} + \beta_t \frac{\partial
}{\partial t}  +  \beta_\sigma \frac{\partial }{\partial \sigma} +
\beta_\rho \frac{\partial }{\partial \rho} \, .
\end{eqnarray}

The Wilson-$\gamma$-functions are easily gathered from the
renormalization factors stated in Eqs.~(\ref{CSEZs}) upon
re-expressing $\mu \frac{\partial }{\partial \mu}$ as $\beta_t
\frac{\partial }{\partial t}$. Because the Wilson functions must
be finite, one then immediately gets
\begin{equation}
\gamma_{\ldots }(t, \sigma ,\rho ) = - t \partial_t X_{\ldots
}^{(1)}(t, \sigma ,\rho ) \, ,
\end{equation}
 where $X_{\ldots }^{(1)}$ is defined in
Eq.~(\ref{expZ}).  Since we will need them to determine the fixed
points of the RG flow, we state the Wilson-$\beta$-functions
explicitly:
\begin{subequations}
\label{betas}
\begin{eqnarray}
\label{betat2}
\beta_t &=& -  t \, \varepsilon +  t^2 \, \frac{20 + 19 \rho -
2\sigma - \rho \sigma}{32\, \pi  \, \sqrt{2+\rho} \, \sqrt{2+\rho
- \sigma} } \, ,
\\
\label{betasigma2}
\beta_\sigma &=&  t \, \frac{\sqrt{2+\rho} \, \sigma \, (3+4\rho -
2\sigma)}{32 \pi  \, \sqrt{2+\rho - \sigma}}\, ,
\\
\label{betarho2}
\beta_\rho &=&  t \, \frac{2 + 7 \rho + 7 \rho^2 + 2 \rho^3}{32
\pi  \, \sqrt{2+\rho} \, \sqrt{2+\rho - \sigma}} \, .
\end{eqnarray}
\end{subequations}

To solve the RG equation we employ the method of characteristics.
We introduce a flow parameter $\ell$ and look for functions
$\bar{\mu} (\ell )$,  $\bar{Z} (\ell )$, $\bar{\omega} (\ell )$,
$\bar{t} (\ell )$, $\bar{\sigma} (\ell )$, and $\bar{\rho} (\ell
)$ determined by the characteristic equations
\begin{subequations}
\begin{eqnarray}
&&\ell \frac{\partial \bar{\mu} (\ell )}{\partial \ell} =
\bar{\mu} \, , \quad \bar{\mu}(1)=\mu \ ,
\\
&&\ell \frac{\partial}{\partial \ell} \ln \bar{Z} (\ell) = \gamma
\left( \bar{t}( \ell) , \bar{\sigma}( \ell) , \bar{\rho}( \ell)
\right) \, , \quad \bar{Z}(1)=1 \,, \qquad
\\
&&\ell \frac{\partial}{\partial \ell} \ln \bar{\omega} (\ell )=
\zeta \left(  \bar{t}( \ell) , \bar{\sigma}( \ell) , \bar{\rho}(
\ell) \right) \, , \quad \bar{\omega}(1)= \omega \, ,
\\
\label{charBetat}
&&\ell \frac{\partial}{\partial \ell}  \bar{t} (\ell ) = \beta_t
\left(  \bar{t}( \ell) , \bar{\sigma}( \ell) , \bar{\rho}( \ell)
\right) \, , \quad \bar{t}(1)= t \, ,
 \\
\label{charBetasigma}
&&\ell \frac{\partial}{\partial \ell}  \bar{\sigma} (\ell ) =
\beta_\sigma \left(  \bar{\sigma}( \ell) , \bar{\sigma}( \ell) ,
\bar{\rho}( \ell) \right) \, , \quad \bar{\sigma}(1)= \sigma \, ,
\qquad
\\
\label{charBetarho}
&&\ell \frac{\partial}{\partial \ell}  \bar{\rho} (\ell ) =
\beta_\rho \left(  \bar{t}( \ell) , \bar{\sigma}( \ell) ,
\bar{\rho}( \ell) \right) \, , \quad \bar{\rho}(1)= \rho \, .
\end{eqnarray}
\end{subequations}
These characteristics describe how the parameters transform if we
change the momentum scale $\mu $ according to $\mu \to
\bar{\mu}(\ell)=\ell \mu $. Being interested in the IR behavior of
the theory, we focus on the limit $\ell \to 0$. In this IR limit
we find that the set of coupling constants $(\bar{t}( \ell) ,
\bar{\sigma}( \ell) , \bar{\rho}( \ell))$ flows to a stable fixed
point
\begin{eqnarray}
(t^\ast , \sigma^\ast , \rho^\ast ) = \left(\frac{32 \, \pi \,
\varepsilon}{7}, 0, -\frac{1}{2} \right)
\end{eqnarray}
satisfying $\beta_t (t^\ast , \sigma^\ast , \rho^\ast ) =
\beta_\sigma (t^\ast , \sigma^\ast , \rho^\ast ) = \beta_\rho
(t^\ast , \sigma^\ast , \rho^\ast ) = 0$. Recalling that $\sigma =
g^2/\omega$ and $\rho = f$, we learn that the stable fixed point
implies two universal ratios of the elastic moduli (Poisson
ratios):
\begin{eqnarray}
C_2^2/(C_1 C_4) = 0 \quad \mbox{and} \quad C_3/C_4 = -1/2\, .
\end{eqnarray}
Note that the longitudinal and the transversal directions are
effectively decoupled at the IR stable fixed point. In addition to
the stable fixed point there are 4 unstable fixed points, viz. the
zero temperature fixed point $t^\ast=0$ as well as $(32 \pi
\varepsilon, 0, -1)$, $(64 \sqrt{2/3} \, \pi \, \varepsilon /13,
1/2, -1/2)$, and $(32 \sqrt{2/3} \, \pi \, \varepsilon , -1/2,
-1)$.

With help of the characteristics the RGE is readily solved, at
least formally:
\begin{eqnarray}
\label{solRG}
&&\Gamma^{(M,N)} \left( \left\{ {\rm{\bf q}}_\perp ,q_d \right\} ;
\omega , t, \sigma , \rho , \mu \right) = \bar{Z}(\ell)^{-(M +
N/2)}
\nonumber  \\
&&\times \,  \Gamma^{(M,N)} \left( \left\{ {\rm{\bf q}}_\perp ,q_d
\right\} ;  \bar{\omega} (\ell), \bar{t}(\ell), \bar{\sigma}(\ell)
, \bar{\rho}(\ell) ,  \mu \ell \right) \, .
\end{eqnarray}
As it stands, Eq.~(\ref{solRG}) does not account for the naive
dimensions of its ingredients. Recalling the $\mu$ invariance of
the Hamiltonian $\mathcal{H}$ it is straightforward to check that
\begin{eqnarray}
\label{dimRG}
&&\Gamma^{(M,N)} \left( \left\{ {\rm{\bf q}}_\perp ,q_d \right\} ;
\omega , t, \sigma , \rho , \mu \right)
\nonumber  \\
&&= \,  \mu^{- (d+1) + dM + (d+1) N}
\nonumber \\
&& \times \, \Gamma^{(M,N)} \left( \left\{ \frac{{\rm{\bf
q}}_\perp}{\mu} ,\frac{q_d}{\mu^2} \right\} ;   \omega , t, \sigma
, \rho , 1 \right) \, .
\end{eqnarray}
Moreover, the $\beta$ invariance tells us that
\begin{eqnarray}
\label{longRG}
&&\Gamma^{(M,N)} \left( \left\{ {\rm{\bf q}}_\perp ,q_d \right\} ;
\omega , t, \sigma , \rho , \mu \right)
 \\
&&= \,  \beta^{(M+N-1)}  \, \Gamma^{(M,N)} \left( \left\{ {\rm{\bf
q}}_\perp ,\frac{q_d}{\beta} \right\} ;   \beta^2 \omega , t,
\sigma , \rho , \mu \right)
\nonumber \\
&&= \,   \omega^{-(M+N-1)/2} \, \Gamma^{(M,N)} \left( \left\{
{\rm{\bf q}}_\perp , \omega^{1/2}q_d \right\} ;   1 , t, \sigma ,
\rho , \mu \right) \, . \nonumber
\end{eqnarray}
where the last line reflects our freedom to chose $\beta =
\omega^{-1/2}$. Combining Eqs.~(\ref{solRG}), (\ref{dimRG}) we
find that the scaling behavior of the vertex functions is
described by
\begin{eqnarray}
\label{totalSol1}
&&\Gamma^{(M,N)} \left( \left\{ {\rm{\bf q}}_\perp ,q_d \right\} ;
\omega , t, \sigma , \rho , \mu \right)
\nonumber \\
&&= \, (\mu  \ell)^{- (d+1) + dM + (d+1) N} \bar{Z}(\ell)^{- (M +
N/2)}
\\
&& \times \, \Gamma^{(M,N)} \left( \left\{ \frac{{\rm{\bf
q}}_\perp}{\mu \ell} ,\frac{q_d}{(\mu\ell)^{2}} \right\} ;
\bar{\omega}(\ell) , \bar{t}(\ell), \bar{\sigma} (\ell) ,
\bar{\rho}(\ell) ,  \mu \ell \right) \, . \nonumber
\end{eqnarray}
Due to Eq.~(\ref{longRG}) we may also write
\begin{eqnarray}
\label{totalSol2}
&&\Gamma^{(M,N)} \left( \left\{ {\rm{\bf q}}_\perp ,q_d \right\} ;
\omega , t, \sigma , \rho , \mu \right)
 \\
&&= \,  (\mu \ell)^{- (d+1) + dM + (d+1) N} \bar{Z}(\ell)^{- (M +
N/2)}  \bar{ \omega}(\ell)^{-(M+N-1)/2}
\nonumber \\
&& \times \, \Gamma^{(M,N)} \left( \left\{ \frac{{\rm{\bf
q}}_\perp}{\mu \ell} ,\frac{\bar{\omega}(\ell)^{1/2} \,
q_d}{(\mu\ell)^{2}} \right\} ;  1 , \bar{t}(\ell), \bar{\sigma}
(\ell) , \bar{\rho}(\ell) ,  \mu \ell \right) \, . \nonumber
\end{eqnarray}
We have to mention that Eqs.~(\ref{totalSol1}) and
(\ref{totalSol2}) do not correctly describe $\Gamma^{(2,0)}_{ab}$
at $\brm{q}_\perp =0$ because we omitted a dangerous irrelevant
bending of the type $K_d \, \brm{q}_d^4$. Without such a term,
Eqs.~(\ref{totalSol1}) and (\ref{totalSol2}) unphysically suggest
that the leading scaling behavior of $\Gamma^{(2,0)}_{ab}$ is
independent of $q_d$ for $\brm{q}_\perp =0$. In other words, $K_d$
is dangerous irrelevant as far as the leading behavior of
$\Gamma^{(2,0)}_{ab}$ at $\brm{q}_\perp =0$ is concerned. Because
$K_d$ is irrelevant its omission has no impact on the leading
behavior of the relevant elastic constants and hence does not
affect our main results. Of course one could and it would be
interesting to investigate the scaling behavior of $K_d$. One has
to keep in mind, however, that irrelevant terms tend to mix under
renormalization with a whole bunch of other irrelevant  terms,
making a proper RG analysis a tedious endevour. This will be left
to future work.

\subsubsection{Scaling 2: Physical quantities}
\label{physCSE}
The variables we considered so far in our RG analysis had the
benefit of being convenient. The flip side of this convenience is,
however, that the featured quantities have no direct physical
meaning. Now we recast our results so that their physical content
becomes pronounced.

To have a clear distinction between physical variables and the
scaled variables we used in our calculations we mark the latter in
the remainder of this section with a hat, i.e., we denote
\begin{eqnarray}
\hat{q}_d = \frac{C_4}{K^2} q_d \, , \ \hat{u}_a = \frac{C_4}{K}
u_a \, , \ \hat{u}_d = \sqrt{\frac{C_4}{K}} u_d \, .
\end{eqnarray}
It is not difficult to see that the relation between the physical
vertex functions and the vertex functions in the scaled variables
is given by
\begin{eqnarray}
\label{rel}
&& \Gamma^{(M,N)} \left( \left\{ {\rm{\bf q}}_\perp ,q_d \right\}
;  \omega , t, \sigma , \rho , \mu \right)
 \\
&& = \, K^{M + 3N/2 -2} C_4^{1-N/2} \, \hat{\Gamma}^{(M,N)} \left(
\left\{ {\rm{\bf q}}_\perp ,\hat{q}_d \right\} ;  \omega , t,
\sigma , \rho , \mu \right) \nonumber  \, .
\end{eqnarray}

Blending Eq.~(\ref{rel}) and our findings of Sec.~\ref{CSERG} we
now obtain
\begin{eqnarray}
\label{physRes}
&&\Gamma^{(M,N)} \left( \left\{ {\rm{\bf q}}_\perp ,q_d \right\} ;
\omega , t, \sigma , \rho , \mu \right) =
\frac{1}{T}\,C_1^{-(M+N-2)/2}
\nonumber \\
&&\times \, C_4^{(2M+N-2)/2} K^{-(M-2)/2} L_\perp^{2(d-1) -dM
-(d+1)N}
\nonumber \\
&&\times \,  \ell^{- (d+1) + dM + (d+1) N}   \left( \bar{
\omega}(\ell)/\omega \right)^{-(M+N-1)/2} \bar{Z}(\ell)^{- (M +
N/2)}
\nonumber \\
&& \times \, \hat{\Phi}^{(M,N)} \bigg( \bigg\{ \frac{ L_\perp
{\rm{\bf q}}_\perp}{\ell} ,\frac{\sqrt{\bar{ \omega}(\ell)/\omega}
L_d q_d}{\ell^{2}} \bigg\} ;
\nonumber \\
&& \hspace{4cm} 1 , \bar{t} (\ell), \bar{\sigma} (\ell) ,
\bar{\rho} (\ell) ,  1 \bigg) \, ,
\end{eqnarray}
where we introduced the susceptibilities
\begin{eqnarray}
\hat{\Phi}^{(M,N)} = t \, \hat{\Gamma}^{(M,N)}
\end{eqnarray}
to make explicit that the vertex functions are proportional to the
inverse temperature. Moreover, we introduced the length scales
\begin{subequations}
\begin{eqnarray}
L_\perp &=& \mu^{-1}  \, ,
\\
\label{defLD}
L_d &=& \frac{C_4}{K^2} \, \frac{\sqrt{\omega}}{\mu^2} =
\sqrt{\frac{C_1}{K}} \, L_\perp^2 \, .
\end{eqnarray}
\end{subequations}
At this stage, $L_\perp$ and $L_d$ are still arbitrary. Further
below we will fix these length scales so that they acquire a
physical meaning, viz.\ the borderline between harmonic and
scaling behavior. Note that the result~(\ref{physRes}) is general
in the sense that it holds in the harmonic as well as in the
scaling limit. When $\bar{t}(l) \approx 0$, the system behaves
approximately like a harmonic system.  When $l$ is small, behavior
is determined by the fixed point with $t^* \sim \epsilon$.

Now to our main goal, viz.\ the behavior of the elastic constants.
The sought after behavior can be inferred without much effort from
result~(\ref{physRes}). As an example, we consider the case $M=0$,
$N=2$ in some detail. In addition to the information contained in
Eq.~(\ref{physRes}) we need some knowledge on the concrete form of
the scaling function $\hat{\Phi}^{(0,2)}_{dd}$. Since $t^\ast$ is
of order $\varepsilon$ it is reasonable to assume that
$\hat{\Phi}^{(0,2)}_{dd}$ can be approximated by its Gaussian form
[cf.\ Eq.~(\ref{GaussVertices})] even in the scaling limit. Hence,
we write\begin{eqnarray}
\label{complikaziert}
&&\hat{\Phi}^{(0,2)}_{dd} \bigg( \bigg\{ \frac{ L_\perp {\rm{\bf
q}}_\perp}{\ell} ,\frac{\sqrt{\bar{\omega}(\ell) /\omega} \, L_d
q_d}{\ell^{2}} \bigg\} ;  1 , \bar{t} (\ell), \bar{\sigma} (\ell)
, \bar{\rho} (\ell) ,  1 \bigg)
\nonumber \\
&&= \, \left( \frac{\sqrt{\bar{\omega}(\ell) /\omega} \, L_d
q_d}{\ell^{2}} \right)^2 + \left(  \frac{L_\perp {\rm{\bf
q}}_\perp}{\ell}\right)^4 \, .
\nonumber \\
\end{eqnarray}
Merging Eqs.~(\ref{complikaziert}) and (\ref{physRes}) we obtain
the physical vertex function
\begin{eqnarray}
\Gamma^{(0,2)}_{dd} \left(  \brm{q}_\perp, q_d  \right) =  T^{-1}
\left\{ C_1 (\ell)\,  q_d^2 + K (\ell ) \,  | \brm{q}_\perp|^{4}
\right\}  \,
\end{eqnarray}
with
\begin{subequations}
\label{constell}
\begin{eqnarray}
\label{resK}
K (\ell ) &=& K\,  \ell^{-\varepsilon} \bar{Z}(\ell)^{-1} \left[
\bar{t}(\ell)/t \right]^{-1} \left[ \bar{ \omega}(\ell)/\omega
\right]^{-1/2} \, ,
\nonumber \\
\\
C_1 (\ell ) &=& C_1 \, \ell^{-\varepsilon} \bar{Z}(\ell)^{-1}
\left[ \bar{t}(\ell)/t \right]^{-1} \left[ \bar{
\omega}(\ell)/\omega \right]^{1/2} \, .
\nonumber\\
\end{eqnarray}
\end{subequations}

In the case of the bending modulus we cannot finalize our
conclusions without solving the characteristics, i.e. without
knowing $\bar{Z}(\ell)$ and so on. For $C_1$, however, we observe
without further information the following:
\begin{eqnarray}
\ell \frac{\partial}{\partial \ell} \, C_1 (\ell ) = - \varepsilon
- \gamma - \frac{\beta_t}{t}  + \frac{\zeta}{2} = - \gamma_\omega
\, .
\end{eqnarray}
Since $\gamma_\omega$ is proportional to $\sigma$ it vanishes at
the stable fixed point. Hence, $C_1$ is independent of $\ell$. In
other words, $C_1$ is normal. Of course, we cannot tell from our
analysis if this stays true beyond 1-loop order. Nevertheless,
this may well be the case.

The behavior of the remaining elastic constants can be extracted
by similar means from the other 2-leg vertex functions. We find
\begin{subequations}
\begin{eqnarray}
\Gamma^{(1,1)}_{ad} \left(  \brm{q}_\perp, q_d  \right) &=& T^{-1}
C_2 (\ell ) \,  q_a q_d  \, ,
\\
\Gamma^{(2,0)}_{ab} \left(  \brm{q}_\perp, q_d  \right) &=& T^{-1}
\big\{ C_3  (\ell ) \, q_a q_b
\nonumber \\
&+&  C_4  (\ell ) ( q_a q_b + \delta_{ab} | \brm{q}_\perp|^{2} )
\big\}  \, .
\end{eqnarray}
\end{subequations}
with the anomalous elastic constants
\begin{subequations}
\label{resConstEll}
\begin{eqnarray}
C_2 (\ell ) &=& C_2\,  \ell^{-\varepsilon} \bar{Z}(\ell)^{-3/2}
\left[ \bar{t}(\ell)/t \right]^{-1} \left[ \bar{
\sigma}(\ell)/\sigma \right]^{1/2}  ,
\\
C_3 (\ell ) &=& C_3 \, \ell^{-\varepsilon} \bar{Z}(\ell)^{-2}
\left[ \bar{t}(\ell)/t \right]^{-1} \left[ \bar{
\omega}(\ell)/\omega \right]^{-1/2}  , \qquad
\\
C_4 (\ell ) &\sim&  C_3 (\ell ) \, .
\end{eqnarray}
\end{subequations}
To obtain the equation for $C_2 (\ell )$ we used the fact that it
must be proportional to $C_2$ and thus to $\sigma^{1/2}$. Note
that $\sigma$ affects the leading behavior of $C_2$ despite
flowing to zero. In other words: $\sigma$ is a dangerous
irrelevant variable. Similar arguments for $C_3$ imply that $C_3
(\ell)$ must be proportional to $\rho$ which reaches a non-zero
fixed point value. There are, therefore, no contributions to the
scaling of $C_3$ from dangerous irrelevant variables.

\subsubsection{Behavior of the elastic constants in $d < 3$}
\label{scaleCSE}
For $\varepsilon >0$ we can assign a physical meaning to the
hitherto arbitrary length scale $L_\perp$ via the definition of
our dimensionless temperature $t$, viz.\
\begin{eqnarray}
\label{physLperp}
L_\perp = \ \left(  \frac{\sqrt{C_1 K^3} \, t}{C_4 \, T}
\right)^{1/\varepsilon} \, .
\end{eqnarray}
Of course, this choice also affects the length scale $L_d$, cf.\
Eq.~(\ref{defLD}). The length scales $L_\perp$ and $L_d$ mark the
borderline between harmonic and critical behavior.

Below 3 dimensions the solutions to the characteristics are for
$\ell \ll 1$ governed by the IR-stable fixed point. Readily, one
finds the power laws
\begin{subequations}
\label{CSEpowerLaws}
\begin{eqnarray}
\bar{Z}(\ell) &=& \ell^{\gamma^\ast} \, ,
\\
\bar{\omega} (\ell) &=& \omega \,  \ell^{\zeta^\ast}\, ,
\end{eqnarray}
where
\begin{eqnarray}
\gamma^\ast & = &\gamma (t^\ast , \sigma^\ast , \rho^\ast ) = - 5
\varepsilon /7 \, ,
\\
\zeta^\ast &= &\zeta (t^\ast , \sigma^\ast , \rho^\ast ) = 4
\varepsilon /7 \, .
\end{eqnarray}
From our discussion above it is clear that we cannot simply set
$\bar{\sigma}(\ell)$ equal to its fixed point value $\sigma^\ast
=0$ because $\sigma$ is a dangerous irrelevant variable that
effects the scaling behavior of $C_2$ at leading order. The
vanishing of $\sigma$ is described by
\begin{eqnarray}
\bar{\sigma}(\ell) \sim \sigma \, \ell^{w}\, ,
\end{eqnarray}
\end{subequations}
with a Wegner exponent $w = \varepsilon/7$ corresponding to the
smallest eigenvalue of the Hessian of the flow near $(t^\ast ,
\sigma^\ast , \rho^\ast )$.

Combining the formal scaling form of Eq.\ (\ref{physRes}) with the
power laws of Eq.\ (\ref{CSEpowerLaws}) for $\bar{Z}(\ell)$,
$\bar{\omega} (\ell)/\omega$ and $\bar{\sigma}(\ell)/ \sigma$ and
the crossover length scales as given in Eqs.~(\ref{physLperp}) and
(\ref{defLD}), we obtain a complete picture of the scaling
behavior of the displacement vertex functions in $d<3$. In
particular we obtain Eqs.~(\ref{CSEscaleForms}) for the 2-point
functions $\Gamma^{(0,2)}_{dd}$, $\Gamma^{(1,1)}_{ad}$ and
$\Gamma^{(2,0)}_{ab}$ with $C_5 =0$ featuring the scaling
exponents
\begin{subequations}
\begin{eqnarray}
&&\eta_K = \varepsilon + \gamma^\ast + \zeta^\ast/2 = 4
\varepsilon /7 \, ,
\\
&&\eta_C = -\varepsilon - 2 \gamma^\ast  -  \zeta^\ast /2 =
\varepsilon/7\, ,
\\
&&\eta_2 = -\varepsilon - 3\gamma^\ast/2 + w/2 =  \varepsilon/7 =
\eta_C\, ,
\\
&&\phi = 2 - \zeta^\ast /2 = 2 - 2 \varepsilon/7 = (4-\eta_K)/2 \,
.
\end{eqnarray}
\end{subequations}
Thus, even though there are three independent exponents
$\gamma^\ast$, $\zeta^\ast$ and $w$ the scaling behavior of the
2-point functions is determined, at least to first order in
$\varepsilon$, by only two exponents, say $\eta_K$ and $\eta_C$.
Upon inserting the power laws~(\ref{CSEpowerLaws}) into
Eqs.~(\ref{resConstEll}) and (\ref{resK}) we obtain the scaling
expressions for $C_2$, $C_3$, $C_4$ and $K$ stated in
Eqs.~(\ref{revCSEC}) and (\ref{revCSEK}) the review section. Note
that $K$ diverges at long length scales whereas $C_2$, $C_3$, and
$C_4$ vanish in this limit.

\subsubsection{Logarithmic behavior in $d=3$}
Since $\varepsilon$ vanishes in $d=3$ the solutions to the
characteristic equations are  no longer of power law type. The
flow of the temperature, for example, is described at leading
order, i.e., for $\bar{\sigma} (\ell) = \sigma^\ast$ and
$\bar{\rho} (\ell) = \rho^\ast$, by
\begin{eqnarray}
\ell \frac{\partial}{\partial \ell}  \bar{t} (\ell ) = \frac{7}{32
\, \pi} \, \bar{t} (\ell )^2 \, .
\end{eqnarray}
This differential equation is readily solved with the result
\begin{eqnarray}
\bar{t}( \ell) /t = \left[1 - \frac{7  \, t}{32 \,  \pi} \ln (\ell
) \right]^{-1} \, .
\end{eqnarray}
Similarly, we find $\bar{Z} (\ell) \sim [\bar{t}( \ell) /t]^{-
\frac{5}{7} }$, $\bar{\omega} (\ell) \sim [\bar{t}( \ell)
/t]^{\frac{4}{7} }$ and $\sigma (\ell) \sim [\bar{t}( \ell) /t]^{
\frac{1}{7}}$ at leading order. Inserting these logarithmic
solutions into Eqs.~(\ref{resK}) and (\ref{resConstEll}) we find
that
\begin{subequations}
\label{log}
\begin{eqnarray}
K (\brm{q}_\perp ) &\sim& K \, \bigg[ 1 - \frac{7  t}{32 \, \pi}
\ln \left(  L_\perp | \brm{q}_\perp| \right) \bigg]^{4/7}  ,
\\
C_2 (\brm{q}_\perp ) &\sim& C_2 \, \bigg[ 1 - \frac{7  \, t}{32 \,
\pi} \ln \left(  L_\perp | \brm{q}_\perp| \right) \bigg]^{-1/7}  ,
\\
C_3 (\brm{q}_\perp ) &\sim& C_4 (\brm{q}_\perp ) \sim C_2
(\brm{q}_\perp ) \, .
\end{eqnarray}
\end{subequations}
Of course, $C_1$ is normal as it was in $d<3$. Once more, $K$
diverges at long length scales and $C_2$, $C_3$, and $C_4$ vanish
in this limit.

Note that Eqs.~(\ref{log}) imply the existence of a nonlinear
crossover length scale, viz.\
\begin{eqnarray}
\xi_\perp = L_\perp \exp \left( \frac{32 \, \pi}{7  \, t} \right)
= L_\perp \exp \left( \frac{32 \, \pi \, \sqrt{C_1 K^3}}{7  \, T
\, C_4} \right)  .\quad
\end{eqnarray}
For $\xi_\perp | \brm{q}_\perp| \approx 1$ the anomalous elastic
constants are approximately harmonic whereas one has clearly
anomalous behavior for $\xi_\perp | \brm{q}_\perp| \ll 1$.

\subsubsection{$C_5$ as a relevant perturbation -- semi-soft elasticity}
\label{pertC5}
Up to this point we have excluded a term proportional to $C_5$,
cf.~Eqs.~(\ref{CSEdefEn}) and (\ref{elasten}), from the elastic
energy of CSEs because such a term destroys the soft elasticity.
Now we establish contact to more conventional uniaxial elastomers
by incorporating a small but nonvanishing $C_5$. As a consequence,
we find semi-soft behavior.

Technically, we treat $C_5$ as a perturbation to the CSE model. It
turns out that this perturbation is relevant in the sense of the
RG. This situation is analogous to the $\phi^4$ model where a
deviation from the critical temperature represents a relevant
perturbation. Our central task we will be to determine the scaling
exponent that governs the departure of $C_5$ from zero.

Our analysis here is based on the full
Hamiltonian~(\ref{CSEdefEn}). Carrying out the $\mu$-rescaling it
is straightforward to see that not the entire strain $u_{ad}$ is
relevant and that it is sufficient to keep
\begin{eqnarray}
\label{truncUad}
u_{ad} = \frac{1}{2} \partial_a u_d\, .
\end{eqnarray}
Applying the rescaling that led us from Eq.~(\ref{Hamil}) to
(\ref{fieldHamil}) we obtain
\begin{eqnarray}
\label{fieldHamile}
\mathcal{H}_e = \mathcal{H} + \frac{e}{2} \int d^{d_\perp} x_\perp
\int d x_d  \,   \partial_a u_d \partial_a u_d \, ,
\end{eqnarray}
with $\mathcal{H}$ as stated in Eq.~(\ref{fieldHamil}) and where
$e = C_5/(4K)$. Note that $e\sim \mu^2$, i.e., the naive dimension
of $e$ is 2 and hence $e$ is clearly relevant.

The Gaussian part of $\mathcal{H}_e$ has an extra term compared to
$\mathcal{H}$ and hence the Gaussian propagator of $\mathcal{H}_e$
is different from that of $\mathcal{H}$. Here, any
$\brm{q}_\perp^4$ in Eqs.~(\ref{GaussProp}) has to be replaced by
$\brm{q}_\perp^4 + e \, \brm{q}_\perp^2$. The non-Gaussian terms
of $\mathcal{H}_e$ and $\mathcal{H}$ are identical and hence we
still have the 4 vertices stated in Eqs.~(\ref{CSEvertices}).

To investigate the departure of $e$ from zero we expand the
propagator to linear order in $e$. Then this expanded propagator
is used in our diagrammatic calculation. Of course, at zeroth
order in $e$ we retrieve our vertex functions~(\ref{vertexFkts}).
The first order in the expansion leads to an extra divergent term
in $\Gamma^{(0,2)}_{dd}$ that is proportional to $e \,
\brm{q}_\perp^2$. To remove the extra divergence we introduce an
additional renormalization factor via setting
\begin{eqnarray}
e \to \mathaccent"7017{e} &=&  Z_T^{-1} Z_e e \, .
\end{eqnarray}
>From the diagrams depicted in Fig.~\ref{diagrams1} we extract that
\begin{eqnarray}
Z_e  &=& 1 -  t \, \frac{6 + 5 \rho + \rho^2}{ 16 \, \pi \,
\varepsilon \, \sqrt{2+\rho} \, \sqrt{2+\rho - \sigma} }  \, .
\end{eqnarray}
The RGE for the vertex functions expanded to linear order in $e$
reads
\begin{eqnarray}
\label{RGEe}
&&\left[ D_{\mu} + e \, \psi \, \partial_e -  \left( M +
\frac{N}{2} \right) \gamma \right]
\nonumber \\
&& \times \, \Gamma^{(M,N)} \left( \left\{ {\rm{\bf q}}_\perp ,q_d
\right\} ;  \omega , t, \sigma , \rho , e, \mu \right) = 0
\end{eqnarray}
with $D_\mu$ as stated in Eq.~(\ref{rgOp}) and
\begin{eqnarray}
\psi = \mu \partial_\mu \ln e |_0 = \gamma_T - \gamma_e\, .
\end{eqnarray}
Setting up a characteristic for $e$,
\begin{eqnarray}
\ell \frac{\partial}{\partial \ell} \ln \bar{e} (\ell )&=& \psi
\left(  \bar{t}( \ell) , \bar{\sigma}( \ell) , \bar{\rho}( \ell)
\right) \, , \quad \bar{e}(1)= e \, , \quad
\end{eqnarray} we find that this coupling flows in $d<3$ as
\begin{eqnarray}
\bar{e} (\ell) \sim e \, \ell^{\psi^\ast} \, ,
\end{eqnarray}
where $\psi^\ast = \psi (t^\ast, \sigma^\ast, \rho^\ast) =
-\varepsilon/7$. Now, the solution to the RGE~(\ref{RGEe}) in
conjunction with dimensional analysis tells us that the scaling
behavior of $\Gamma^{(0,2)}_{dd}$ expanded to linear order in $e$
is given by
\begin{eqnarray}
\label{totalSole}
&&\Gamma^{(0,2)}_{dd} \left(  {\rm{\bf q}}_\perp ,q_d ;  \omega ,
t, \sigma , \rho , e, \mu \right) =  (\mu \ell)^{4 -\varepsilon}
\bar{Z} (\ell)^{-1}
\\
&&\times \, \Gamma^{(0,2)}_{dd} \left(  \frac{{\rm{\bf q}}_\perp}{
\mu \ell} , \frac{q_d}{(\mu \ell)^2} ;  \bar{\omega}(\ell) ,
\bar{t}(\ell), \bar{\sigma} (\ell) , \bar{\rho}(\ell) ,
\frac{\bar{e} (\ell)}{(\mu \ell)^2}, 1 \right) \, . \nonumber
\end{eqnarray}
Next we switch back to physical variables. By performing much the
same steps as in Secs.~(\ref{scaleCSE}) and (\ref{physCSE}) we
obtain for $d<3$
\begin{eqnarray}
\label{physSole}
&&\Gamma^{(0,2)}_{dd} \left(  {\rm{\bf q}}_\perp ,q_d ;  e \right)
= \frac{K}{T} \, L_\perp^{-4} \ell^{4- \eta_K}
\\
&&\times \, \hat{\Phi}^{(0,2)}_{dd} \left(  \frac{L_\perp{\rm{\bf
q}}_\perp}{ \ell} , \frac{L_d q_d}{\ell^\phi} ;   \frac{L_\perp^2
\, e}{\ell^{1/\nu_5}} \right) \, , \nonumber
\end{eqnarray}
where we have dropped several arguments for notational simplicity
and where
\begin{eqnarray}
\nu_5= (2 -\psi^\ast)^{-1}= 1/2 - \varepsilon/28 \, .
\end{eqnarray}

Thus, $C_5$ plays the same role in this problem as temperature
plays in a traditional thermal phase transition. As in the thermal
case, it is useful to introduce a correlation length
\begin{eqnarray}
\xi_5 = L_\perp (e L_\perp^2)^{-\nu_5} \sim C_5^{-\nu_5} \, .
\end{eqnarray}
There are two interesting limits we can now consider: $\xi_5
|{\rm{\bf q}}_\perp| \ll 1$ and $\xi_5 |{\rm{\bf q}}_\perp| \gg
1$. In the first case, $\Gamma^{(0,2)}_{dd}$ must be proportional
to $\brm{q}_\perp^2$ when $q_d =0$ and to $q_d^2$ when
$\brm{q}_\perp = \brm{0}$. Thus, we obtain by choosing $\ell
=(\xi_5/L_\perp)^{-\nu_5}$ from Eq.~(\ref{physSole}):
\begin{eqnarray}
&&\Gamma^{(0,2)}_{dd} \left(  {\rm{\bf q}}_\perp ,q_d ;  e \right)
\\
&& \sim \frac{1}{T}  \left\{
\begin{array}{lcl}
K L_\perp^{-\eta_K} \xi_5^{-(2-\eta_K)}  \brm{q}_\perp^2 \sim
C_5^{\gamma_5} \brm{q}_\perp^2   & \quad \mbox{if}  & \quad  q_d
=0
\\
C_1 q_d^2  &\quad \mbox{if} &\quad \brm{q}_\perp = \brm{0}
\end{array}
\right. \nonumber
\end{eqnarray}
where
\begin{eqnarray}
\gamma_5 = \nu_5\, (2-\eta_K) = 1 -5\, \varepsilon/14 \, .
\end{eqnarray}
The second limit $\xi_5 |{\rm{\bf q}}_\perp| \gg 1$ corresponds
$C_5 \to 0$, and we must obtain the same scaling forms as we
obtained for $C_5 =0$ with a correction term that vanishes with
$(\xi_5 |{\rm{\bf q}}_\perp|)^{-1/\nu_5}$.

As we stated earlier, we have not included nonlinear terms in our
model Hamiltonian [Eq.\ (\ref{Hamil})] that are needed to
stabilize the system when $C_5<0$.  In particular, we have not
included terms $(u_{ad} u_{ad})^2 \sim (\partial_a u_d \partial_a
u_d)^2$, $u_{aa}u_{bd} u_{bd}$, $u_{ab} u_{ad}u_{bd}$, $u_{aa}^2
u_{bd} u_{bd}$, $u_{ab}^2 u_{cd}u_{cd}$, and $u_{aa} u_{bc} u_{bd}
u_{cd}$, all of which have the same naive scaling dimension as the
harmonic terms in nonlinear strain that we retained in Eq.\
(\ref{Hamil}).  Our expectation is that the general theory in
which these terms are included will have the same general form as
the present theory with, however, a different stable fixed point
with in particular a nonzero value of the coefficient of
$(u_{ad}u_{ad})^2$.  At such a fixed point, $\langle \partial_a
u_d \rangle$ will develop a nonzero value at negative $C_5$ that
scales as $(-C_5)^{\beta}$ where $\beta = (1 + \case{1}{4}
\gamma^*) \nu_5$.  This result can be obtained by observing that
Eq.\ (\ref{physRes}) for $\Gamma^{(0,2)}$ implies  $\partial_a^2
G_{dd}(x_{d}, x_{a}, C_5) = l^{2 + \gamma^*/2}\partial_a^{\prime
2} G_{dd} ( l^{2 + \zeta^*/2} x_d, l x_a, l^{-\nu_5} C_5)$, where
$\partial_a^{\prime}$ is a derivative with respect to $l x_a$, and
that $G_{dd}(x_{d}, x_{a}, C_5)\rightarrow \langle
\partial_a u_d\rangle \langle \partial_a u_d\rangle$ as
$\brm{x}\rightarrow \infty$.

\section{Nematic Elastomers}
\label{ne}
Usual NEs are either crosslinked in the isotropic or in the
nematic phase. If synthesized in the isotropic phase, the uniaxial
anisotropy arises via a spontaneous symmetry breaking at the
isotropic to nematic transition and is associated with soft
elasticity. Crosslinking in the nematic phase, on the other hand,
permanently imprints the uniaxial anisotropy into the material and
leads to semi-soft behavior. We will start by studying the soft
case. Further below, we will include the effects of an imprinted
uniaxial anisotropy to investigate the semi-soft case.

\subsection{The model}
\label{neModel}
Since the spontaneous symmetry axis of soft NEs can point in any
direction, their elastic energy has to be rotationally invariant
not only in target space but also in  reference space. Both
invariances are taken into account by writing the stretching
energy as
\begin{eqnarray}
\label{originalNEenergy}
&&H_{\text{st}} =  \int d^d x  \, \Big\{  \frac{\lambda}{2} \big(
\tr \tens{u}  \big)^2 + \mu   \tr \tens{u}^2 + A_1 \big(  \tr
\tens{u} \big)^3
\nonumber \\
&& + \, A_2 \tr \tens{u}  \tr \tens{u}^2  + A_3 \tr \tens{u}^3 +
B_1 \big(  \tr \tens{u} \big)^4 + B_2 \big(  \tr \tens{u}^2
\big)^2
\nonumber \\
&& + \, B_3 \big(  \tr \tens{u} \big)^2 \tr \tens{u}^2 + B_4 \tr
\tens{u}  \tr \tens{u}^3 +  B_5  \tr \tens{u}^4  \Big\}\, ,
\end{eqnarray}
with the first two expansion coefficients being the usual Lam\'{e}
coefficients. Of course, terms of higher than fourth order are
allowed by the symmetries of the system. However, these
higher-order terms turn out to be irrelevant in the RG sense and
are hence neglected.

Suppose that the spontaneously uniaxially ordered elastomer is
described in equilibrium by an equilibrium strain tensor
$\tens{u}_0$. Without loss of generality we may assume that the
anisotropy axis lies in the $\hat{\brm{e}}_d$ direction and that
$\tens{u}_0$ is a diagonal matrix with the diagonal elements
$u_{0aa} = u_{0\perp}$ and $u_{0dd} = u_{0\parallel}$.  To
describe deviations from the equilibrium configuration, we
introduce the relative strain
\begin{eqnarray}
\label{relStrain}
\tens{w} = \tens{u} - \tens{u}_0
\end{eqnarray}
 and to expand $\mathcal{H}$ in terms of $\tens{w}$. By dropping terms that depend only on $\tens{u}_0$ we find
\begin{eqnarray}
\label{expansionHst1}
&&H_{\text{st}} =   \int d^{d_\perp} x_\perp \int d x_d  \, \big\{
a_1 \, w_{dd} + a_2 \, w_{aa} + b_1 \, w_{dd}^2
\nonumber \\
&&+ \, b_2 \, w_{dd} w_{aa} +  b_3 \, w_{aa}^2 + b_4 \, w_{ab}^2 +
b_5 \, w_{ad}^2 + c_1 \, w_{dd} w_{ad}^2
\nonumber \\
&& +\,  c_2 \, w_{aa} w_{bd}^2 +  c_3 \, w_{ab} w_{ad} w_{bd}  +
d_1 \, w_{ad}^2 w_{bd}^2 \big\} \, ,
\end{eqnarray}
where we have discarded terms that turn out to be irrelevant. The
new coefficients $a_1$, $a_2$, $b_1$, and so on, depend on the old
coefficients $\lambda$, $\mu$, and so forth, as well as on
$u_{0\parallel}$ and $u_{0\perp}$.

By virtue of the rotational invariance in reference space there
exists a set Ward identities relating the vertex functions
implicit in (\ref{expansionHst1}). We derive these identities in
Appendix~\ref{app:ward2}. At zero-loop or mean-field level these
Ward identities correspond to relations between the elastic
constants in (\ref{expansionHst1}),
\begin{subequations}
\label{coeffRels}
\begin{eqnarray}
\label{coeffRelA}
&& a_1 -  a_2 - s \, b_5 = 0 \, ,
\\
&&b_2 - 2 b_3 - s \, c_2 = 0 \, , \quad b_2 + b_5 - 2 b_1 + s\,
c_1 = 0 \, ,\qquad
\\
&&b_5 - 2 b_4 - s \, c_3 = 0 \, , \quad c_1 - c_2 - c_3 - 2 s \,
d_1= 0 \, ,
\end{eqnarray}
\end{subequations}
where $s$ is an abbreviation for $u_{0\parallel} - u_{0\perp}$.
Since $\tens{w}$ describes deviations from the equilibrium
$\tens{u}_0$ its thermal average $\langle \tens{w} \rangle$ has to
vanish. At zero temperature, where the mean-field approximation
becomes exact, this means that the coefficients of the linear
terms in (\ref{expansionHst1}) must be zero.
Equation~(\ref{coeffRelA}) then leads to the observation that $b_5
=0$ for $u_{0\parallel} \neq u_{0\perp}$. At finite temperatures
thermal fluctuations become important and loop corrections
renormalize the elastic constants including $a_1$, $a_2$ and
$b_5$. For $\langle \tens{w} \rangle$ to vanish the renormalized
versions of $a_1$ and $a_2$ have to satisfy equations of state to
which $a_1 = 0$ and $a_2 =0$ are the mean-field approximations. In
the following we will assume that we have chosen $a_1$ and $a_2$
appropriately so that their respective equations of state are
satisfied. In other words: we assume that we expand about the true
equilibrium state. Then, the Ward identity~(\ref{ex3})
generalizing (\ref{coeffRelA}) guarantees that the renormalized
$b_5$ vanishes for $u_{0\parallel} \neq u_{0\perp}$. The vanishing
of the elastic constant $b_5$ is the origin of the softness of NEs
as can easily be seen by rewriting the nonvanishing terms of the
Hamiltonian at leading order in small deformations in Fourier
space.

In what follows  we assume that $a_1 = a_2 = b_5 = 0$. Exploiting
the relations~(\ref{coeffRels}) we rewrite the stretching energy
as
\begin{eqnarray}
\label{expansionHst2}
H_{\text{st}} &=&   \int d^{d_\perp} x_\perp \int d x_d  \, \big\{
b_1 \, v_{dd}^2 + \, b_2 \, v_{dd} v_{aa} +  b_3 \, v_{aa}^2
\nonumber \\
&+& b_4 \, v_{ab}^2  \big\} \, ,
\end{eqnarray}
where we have introduced the non-standard strains
\begin{subequations}
\label{weirdStrains}
\begin{eqnarray}
v_{ab} &=& w_{ab} - s^{-1} w_{ad} w_{bd}\, ,
\\
v_{dd} &=& w_{dd} + s^{-1} w_{ad} w_{ad}\, .
\end{eqnarray}
\end{subequations}
Next we cast $v_{ab}$ and $v_{dd}$ into a more familiar form.
Recall that the relative strain $\tens{w}$ depends upon the
displacement $\brm{u}$ relative to the original isotropic state
measured in the original reference space coordinate $\brm{x}$. It
is more convenient, however, to work with a relative strain
$\tens{u}^\prime$ that depends on the displacement
$\brm{u}^\prime$ relative to the equilibrium state of the NE
measured in the coordinate $\brm{x}^\prime$ of the corresponding
uniaxial reference state. The tensors $\tens{w}$ and
$\tens{u}^\prime$ are related via (see, e.g.,
Ref.~\cite{Lubensky&Co_2001})
\begin{eqnarray}
\label{relTens}
\tens{w} = \tens{\Lambda}_0^T \tens{u}^\prime \tens{\Lambda}_0 \,
,
\end{eqnarray}
where $\tens{\Lambda}_0$ is the so-called Cauchy deformation
tensor of the uniaxial equilibrium state. $\tens{\Lambda}_0$ and
$\tens{u}_0$ provide equivalent descriptions of this state and
they are related via
\begin{eqnarray}
\tens{u}_0 = \frac{1}{2} \left( \tens{\Lambda}_0^T
\tens{\Lambda}_0 - \tens{1} \right) \, ,
\end{eqnarray}
where $\tens{1}$ denotes the $d \times d$ unit matrix.
Substituting the relation~(\ref{relTens}) into
Eqs.~(\ref{weirdStrains}) we obtain
\begin{subequations}
\label{fastda}
\begin{eqnarray}
v_{ab} &=&  \frac{\Lambda_{0\perp}^2}{2}  \left[ \partial_a^\prime
u_b^\prime + \partial_b^\prime  u_a^\prime  - \frac{1}{r-1} \,
\partial_a^\prime  u_d^\prime \partial_a^\prime  u_d^\prime
\right]  , \quad
\\
v_{dd} &=& r \, \Lambda_{0\perp}^2 \left[  \partial_d^\prime
u_d^\prime  +  \frac{1}{2} \, \frac{1}{r-1} \,   \partial_a^\prime
u_d^\prime \partial_a^\prime  u_d^\prime\right]  ,
\end{eqnarray}
\end{subequations}
where we have used Eqs.~(\ref{truncStrains}) and (\ref{truncUad})
to express $\tens{u}^\prime$. $r = \Lambda_{0\parallel}^2
/\Lambda_{0\perp}^2$ is the usual anisotropy ratio that
characterizes the anisotropy of the uniaxial equilibrium state. In
the steps leading to Eqs.~(\ref{fastda}) we have exploited that $s
= \Lambda_{0\perp}^2 (r-1)/2$. A glance at Eqs.~(\ref{fastda})
shows that we can write $\tens{v}$ in a simpler and more
traditional form by rescaling $x_a^\prime \to x_a$, $x_d^\prime
\to \sqrt{r-1} \, x_d$, $u_a^\prime \to u_a$ and $u_d^\prime \to
\sqrt{r-1} \, u_d$. Incorporating bending, we eventually arrive at
the Hamiltonian
\begin{eqnarray}
\label{HamilNE}
\mathcal{H} &=& \frac{1}{2} \int d^{d_\perp} x_\perp \int d x_d \,
\big\{  C_1\,  v_{dd}^2 + K \left(  \nabla_\perp^2 u_d \right)^2
\nonumber \\
&+& 2 \, C_2 \, v_{dd} v_{aa} +  C_3 \, v_{aa}^2 + 2 \, C_4 \,
v_{ab}^2  \big\}  \, ,
\end{eqnarray}
with the non-standard strains
\begin{subequations}
\begin{eqnarray}
v_{ab} &=& \frac{1}{2} (  \partial_a  u_b +  \partial_b  u_a -
\partial_a  u_d \partial_b  u_d ) \, ,
\\
v_{dd} &=&  \partial_d  u_d + \frac{1}{2}  \partial_a  u_d
\partial_b  u_d \, .
\end{eqnarray}
\end{subequations}
Here, we exclusively retained the only bending term that is
relevant in the RG sense. The new elastic constants $C_{...}$ are
proportional to the $b_{...}$ in (\ref{expansionHst2}):$C_1 = 2 \,
b_1 \Lambda_{0\perp}^4 r^2 \sqrt{r-1}$, $C_2 = b_2
\Lambda_{0\perp}^4 r \sqrt{r-1}$, $C_3 = 2 \, b_3
\Lambda_{0\perp}^4 \sqrt{r-1}$ and $C_4 =  b_4 \Lambda_{0\perp}^4
\sqrt{r-1}$.

A comment regarding the rescaling of $x_d$ is called for.  Our
rescalings $x_d^\prime \to \,  \sqrt{r-1}\, x_d$ and $u_d^\prime
\to \sqrt{r-1} \, u_d$ only make sense if $r>1$.  We have assumed
that the nematic phase is characterized by $r>1$.  Nematic phases
with $r<1$ are also possible.  In this case, our rescalings would
be different.  At a second-order phase transition, $r-1$ will tend
to zero.  The transition from the isotropic to the nematic
elastomer phase is first order in all dimensions above 2.  The
order of the transition in two dimensions has not yet been
established.  A second-order transition would have unusual
properties since the coefficients of the nonlinear contributions
to the nonlinear strain of Eqs.~(\ref{fastda}) diverge as $r
\rightarrow 0$.

Instead of using the Hamiltonian~(\ref{Hamil}) or our RG analysis
we find it convenient to reduce the number of constants featured
in the statistical weight $\exp (  - \mathcal{H}/T)$ by rescaling
$T \to \sqrt{K^3 /C_4} T$, $x_d \to \sqrt{C_4 /K}x_d$, $u_d \to
\sqrt{K /C_4} u_d$ and $u_a \to (K /C_4) u_a$. This gives us
finally
\begin{eqnarray}
\label{fieldHamilNE}
\frac{\mathcal{H}}{T} &=& \frac{1}{2\, T} \int d^{d_\perp} x_\perp
\int d x_d  \, \big\{  \omega\,  v_{dd}^2 +  \left( \nabla_\perp^2
u_d \right)^2
\nonumber \\
&+& 2 \, g \, v_{dd} v_{aa} +  f \, v_{aa}^2 + 2  \, v_{ab}^2
\big\}  \, ,
\end{eqnarray}
where
\begin{equation}
\omega = C_1 / C_4, \qquad g = C_2 / C_4 , \qquad f = C_3 / C_4 .
\end{equation}
Like the parametrization of Eq.\ (\ref{fieldHamil}) of the
Hamiltonian for CSEs, the parametrization of
Eq.~(\ref{fieldHamilNE}) is not appropriate for taking the
$C_4\rightarrow \infty$ limit to obtain a smectic-$A$ Hamiltonian.
As was the case for CSEs, we are interested in properties unique
to NEs, and we will not consider the Grinstein-Pelcovits limit of
our model.

Formally, the Hamiltonians~(\ref{fieldHamil}) and
(\ref{fieldHamilNE}) look very similar. One has to bear in mind,
however, that the strains $\tens{u}$ and $\tens{w}$ are different.
In fact, the scaling symmetries of (\ref{fieldHamil}) and
(\ref{fieldHamilNE}) are quite different except for the $\mu$
invariance. The naive dimensions of the fields, the temperature
and the coupling constants are the same for the two models. In
particular, NEs and CSEs have a mutual upper critical dimension
$d_c =3$. Though both systems possess of a mixing invariance the
specific forms of these invariances are distinct: the NE
Hamiltonian is invariant under the transformation $u_a (x_c , x_d)
\to u_a (x_c - \theta_c x_d , x_d) + \theta_a u_d (x_c , x_d)$ and
$u_d  (x_c , x_d) \to u_d (x_c - \theta_c x_d , x_d) + \theta_a
x_a$ provided that the $\theta$'s are small. This mixing
invariance is reminiscent of the original reference space rotation
symmetry. In our current approach to NEs, the $\beta$-invariance of CSEs has counterpart.
However, in Appendix~\ref{app:alternativeScheme} we present an alternative formulation
that features a $\beta$-invariance at the cost of having an extra scaling parameter. Due to the different forms of the strains, the remaining symmetries
stated at the end of Sec.~\ref{iseModel} have no analog in NEs.

\subsection{Renormalization group analysis}
\label{nerg}
The diagrammatic perturbation expansions for CSEs and NEs are
similar. Instead of repeating the details we will high-light the
differences. In contrast, the RG behavior of CSEs and NEs is quite
different due to the varying scaling symmetries.

\subsubsection{Diagrammatic expansion}
It is not difficult to see that the Gaussian propagator for NEs
coincides with that for CSEs, see Eqs.~(\ref{GaussProp}). The
differences between $\tens{u}$ and $\tens{v}$, however, lead to
different vertices. For NEs we have
\begin{subequations}
\label{NEvertices}
\begin{eqnarray}
&&i \, \frac{\omega - g}{2 \, T} \, q_d^{(1)}  q_b^{(2)} q_b^{(3)}
\, ,
\\
&&i \, \frac{g - f}{2 \, T} q_a^{(1)} \,  q_b^{(2)} q_b^{(3)}  \,
,
\\
&&i \,  \frac{1}{T} \, q_a^{(2)}  q_b^{(1)} q_b^{(3)}   \, ,
\\
&&- \, \frac{\omega - 2g + f+2}{8 \, T} \, q_a^{(1)}  q_a^{(2)}
q_b^{(3)}  q_b^{(4)}   \, .
\end{eqnarray}
\end{subequations}
Of course the sum of the momenta has to vanish at each vertex.
Note that the vertices~(\ref{CSEvertices}) and (\ref{NEvertices})
are of the same structure. Merely the coupling constants appear in
different combination. Hence, the Feynman diagrams for both models
have the same topology, or in other words, the 2-leg diagrams for
both models can be drawn as in Figs.~\ref{diagrams1} to
\ref{diagrams3}. Moreover, the same type of vertex functions are
superficially divergent. By virtue of the NE mixing invariance,
there exists a set of Ward identities relating the NE vertex
functions. These identities are derived and stated in
Appendix~\ref{app:ward3}. They guarantee that we merely have to
calculate the 2-point functions. Appendix~\ref{app:diagramCalc}
describes details if the reader makes the appropriate changes in
the vertices. Our results for the 2-point functions read
\begin{subequations}
\label{NEvertexFkts}
\begin{eqnarray}
&&\Gamma^{(0,2)}_{dd} = T^{-1} \big[ \tau +\omega q_d^2 +
\brm{q}_\perp^4 \big]
 \\
& &-\,  \frac{{\sqrt{2 + f}}\,(g-\omega)^2\, q_d^2 }
  {16\,\pi \,\varepsilon \,{\sqrt{-g^2 + \left( 2 + f \right) \omega }}} \, \tau^{-\varepsilon /4 }\, ,
  \nonumber
  \\
&& -\, \frac{\left[ 4g + g^2 - 2(6+\omega) - f (12 + \omega)
\right] \, \brm{q}_\perp^4}{32\,\pi \,\varepsilon \,
    {\sqrt{2 + f}} \, \sqrt{ -g^2 + \left( 2 + f \right) \omega }} \, \tau^{-\varepsilon /4 }\, ,
\nonumber \\
&&\Gamma^{(1,1)}_{ad} = T^{-1} g \, q_a q_d
\\
&& -\, \frac{ {\sqrt{2 + f}} \, \left( 1 + f - g\right) \,
(g-\omega)\,q_d\,
      {q_a} }{16\,\pi \,\varepsilon \,
    {\sqrt{-g^2 + \left( 2 + f \right) \omega }}} \, \tau^{-\varepsilon /4 } \, ,
\nonumber
\\
&&\Gamma^{(2,0)}_{ab} = T^{-1} \big[ (f+1) \, q_a q_b +
\delta_{ab} \, \brm{q}_\perp^2 \big]
 \\
&&-\, \frac{ {\sqrt{2 + f}}\,[ 2\,{\left( 1 + f - g\right)
}^2\,{q_a}
         {q_b} + {{\delta }_{ab}} \, \brm{q}_\perp^2]   }{32\,\pi \,
    \varepsilon \,{\sqrt{-g^2 + \left( 2 + f \right) \omega }}} \,  \tau^{-\varepsilon /4 } \, .
    \nonumber
\end{eqnarray}
\end{subequations}

\subsubsection{Renormalization}
We eliminate the $\varepsilon$ poles from the NE vertex functions
by employing the renormalization scheme
\begin{subequations}
\label{NEscheme}
\begin{eqnarray}
&&x_d \to \mathaccent"7017{x}_d  = Z^{-1/2} x_d \, ,
\\
&&u_d \to \mathaccent"7017{u}_d  = Z^{1/2} u_d \, ,
\\
&&u_a \to \mathaccent"7017{u}_a  = Z u_a \, ,
\\
&&T \to \mathaccent"7017{T} = Z^{1/2} Z^{-1}_T \mu^\varepsilon t
\, ,
\\
&&\omega \to \mathaccent"7017{\omega} =  Z^{-1} Z_T^{-1} Z_\omega
\omega \, ,
\\
&&g \to  \mathaccent"7017{g} = Z^{-1}  Z_T^{-1} Z_g g \, ,
\\
&&f \to  \mathaccent"7017{f} = Z^{-1}  Z_T^{-1} Z_f f \, .
\end{eqnarray}
\end{subequations}
Our renormalizations are devised so that the strains $v_{ab}$ and $v_{dd}$ as well as our Hamiltonian remain invariant in form,
\begin{eqnarray}
\label{renoHamilNE}
&&\frac{\mathcal{H}}{T}  \to \frac{1}{2 \, T} \int d^{d_\perp}
x_\perp \int d x_d  \, \Big\{   Z_\omega \omega \, v_{dd}^2 +  Z_T
\left(  \nabla_\perp^2 u_d \right)^2
\nonumber \\
&&   + \, 2 Z_g g \,  v_{dd} u_{aa}    +  Z_f f \, v_{aa}^2 + 2 \,
Z_T Z v_{ab} v_{ab} \Big\} \, ,
\end{eqnarray}
The scheme~(\ref{NEscheme}) follows closely the approach developed by Grinstein and Pelcovitz~\cite{grinstein_pelcovits_81_82}. Of course, other re-parameteriztions are conceivable. In Appendix~\ref{app:alternativeScheme} we present an alternative formulation with a different renormalization scheme in which neither the elastic displacement nor $x_d$ is renormalized.

In our current formulation, there remanis no scaling invariance of the Hamiltonian that can be exploited to further reduce the number of coupling constants. Hence, the
renormalization factors are functions of the original
dimensionless parameters $t$, $\omega$, $g$ and $f$ rather than of
a reduced number of effective couplings. The structure of the NE
renormalization factors is
\begin{equation}
\label{expZNE}
Z_{\ldots }(t, \omega , g, f)  =1+\sum_{m=1}^{\infty
}\frac{X_{\ldots }^{(m)}(t, \omega , g, f)}{\varepsilon^{m}}\, .
\end{equation}
with $X_{\ldots }^{(m)}(t, \omega , g, f)$ being a power series in
the effective temperature $t$ beginning with the power $t^{m}$. To
1-loop order we find from (\ref{NEvertexFkts}) via minimal
subtraction
\begin{subequations}
\begin{eqnarray}
&&Z  = 1 +  t \, \frac{4g + g^2 - 2(7+\omega) - f (13 +
\omega)}{32\,\pi \,\varepsilon \,
    {\sqrt{2 + f}} \, \sqrt{ -g^2 + \left( 2 + f \right) \omega }} \, ,
\\
&&Z_T =  1 - t\, \frac{4g + g^2 - 2(6+\omega) - f (12 +
\omega)}{32\,\pi \,\varepsilon \,
    {\sqrt{2 + f}} \, \sqrt{ -g^2 + \left( 2 + f \right) \omega }} \, , \qquad
\\
&&Z_\omega  = 1 + t \, \frac{\sqrt{2+f} \, (g-\omega)^2}{16 \, \pi
\, \varepsilon \, \omega \, \sqrt{-g^2 +(2+f)\omega}} \, ,
\\
&&Z_g = 1 + t \, \frac{\sqrt{2+f} \, (1+f-g)\, (g-\omega)}{16\,
\pi \, \varepsilon \, g \, \sqrt{-g^2 +(2+f)\omega}}  \, ,
\\
&&Z_f = 1 + t \, \frac{\sqrt{2+f} \, [ 1+4 (f - g) + 2
(f-g)^2]}{32 \, \pi \, \varepsilon \, f \sqrt{-g^2 +(2+f)\omega}}
\, .\qquad
\nonumber \\
\end{eqnarray}
\end{subequations}

\subsubsection{Scaling 1: RG equation and its solution}
The RGE for the NE vertex functions follows as ususal from the
fact that the unrenormalized theory has to be independent of the
arbitrary length scale $\mu^{-1}$ introduced by renormalization.
Instead of working with the original parameters $\omega$, $g$ and
$f$ we prefer to switch to
\begin{subequations}
\label{gutmuetig}
\begin{eqnarray}
&&\kappa = g/\omega = C_2/C_1\, ,  \quad \rho = f/\omega =
C_3/C_1\, , \quad
\\
& &\sigma = 1/\omega = C_4/C_1 \, .
\end{eqnarray}
\end{subequations}
This step turns out to be helpful in studying the RG flow because
some of the original parameters tend to flow to infinity. We will
see shortly that, on the other hand, $\kappa$, $\rho$ and $\sigma$
flow to finite values. Our RGE reads
\begin{eqnarray}
\left[ D_\mu - \left(  M + \frac{N}{2} \right) \gamma \right]
\Gamma^{(M,N)} \left(  \left\{  \brm{q}_\perp, q_d \right\} ; t,
\kappa , \rho , \sigma, \mu  \right) = 0
\nonumber \\
\end{eqnarray}
with the RG differential operator
\begin{eqnarray}
\label{diffOpNE}
D_\mu = \mu \partial_\mu - \frac{\gamma}{2} q_d \partial_{q_d} +
\beta_t \partial_t +  \beta_\kappa \partial_\kappa +  \beta_\rho
\partial_\rho +  \beta_\sigma \partial_\sigma \, .
\nonumber \\
\end{eqnarray}
The Wilson-$\beta$-functions, from which we determine the fixed
points of the RG flow, are given in terms of the
Wilson-$\gamma$-functions
\begin{eqnarray}
\gamma_{\ldots }  = \mu \frac{\partial }{\partial \mu} \ln
Z_{\ldots } \Big|_0  \, .
\end{eqnarray}
by
\begin{subequations}
\begin{eqnarray}
\beta_t &=&  t\, (-\varepsilon + \gamma_T - \gamma/2) \, ,
\\
\beta_\kappa &=&  \kappa \, (\gamma_\omega - \gamma_g)  \, ,
\\
\beta_\rho &=& \rho \, (\gamma_\omega - \gamma_f) \, ,
 \\
\beta_\sigma &=&  \sigma \, (\gamma_\omega - \gamma_T -\gamma)  \,
.
\end{eqnarray}
\end{subequations}
The Wilson-$\gamma$-functions are readily extracted from the
renormalization factors with the result
\begin{eqnarray}
\gamma_{\ldots} = - t \partial_t X_{\ldots }^{(1)}(t, \omega , g,
f) \, .
\end{eqnarray}
Switching to the parameters defined in Eqs.~(\ref{gutmuetig}) we
obtain
\begin{subequations}
\begin{eqnarray}
\beta_t &=&  - t \, \varepsilon
\\
&-&t^2 \,  \frac{ 3 \kappa^2 + 12\kappa\sigma - 2
\sigma(3+19\sigma)- \rho (3+37\sigma) }{ 64 \, \pi \, \sigma \,
\sqrt{2+\rho/\sigma} \, \sqrt{\rho + 2 \sigma - \kappa^2}} \, ,
\nonumber
\\
\beta_\kappa &=& t \, \frac{\sqrt{2+\rho/\sigma} \, (\kappa -1) \,
(\rho + \sigma - \kappa^2)} {16 \, \pi \,  \sqrt{\rho + 2 \sigma -
\kappa^2}}  \, ,
\\
\beta_\rho &=& t \, \frac{\sqrt{2+\rho/\sigma} }{32 \, \pi \,
\sqrt{\rho + 2 \sigma - \kappa^2}} \, ,
 \\
&\times& \left\{ 2\kappa^2 (1-\rho) + 2\rho^2 - 4 \kappa \sigma +
\sigma^2 + \rho (4\sigma -2)\right\}
\nonumber \\
\beta_\sigma &=& t \, \frac{\sqrt{2+\rho/\sigma} \,\sigma \,  ( -2
+ 4 \kappa - 2\kappa^2 + \sigma)}{32 \, \pi \,  \sqrt{\rho + 2
\sigma - \kappa^2}}  \, .
\end{eqnarray}
\end{subequations}

For solving the RGE we introduce the characteristics
\begin{subequations}
\begin{eqnarray}
&&\ell \frac{\partial \bar{\mu} (\ell )}{\partial \ell} =
\bar{\mu} \, , \quad \bar{\mu}(1)=\mu \ ,
\\
&&\ell \frac{\partial}{\partial \ell} \ln \bar{Z} (\ell)= \gamma
\left(  \bar{t}( \ell) , \bar{\kappa} (\ell ), \bar{\sigma}( \ell)
, \bar{\rho}( \ell) \right) \, , \quad  \bar{Z}(1)=1 \,
\nonumber \\
\\
&&\ell \frac{\partial}{\partial \ell}  \bar{t} (\ell ) = \beta_t
\left(  \bar{t}( \ell) , \bar{\kappa} (\ell ), \bar{\sigma}( \ell)
, \bar{\rho}( \ell) \right) \, , \quad \bar{t}(1)= t \, ,
\nonumber \\
\\
&&\ell \frac{\partial}{\partial \ell}  \bar{\kappa} (\ell )=
\beta_\kappa \left(  \bar{t}( \ell) , \bar{\kappa} (\ell ),
\bar{\sigma}( \ell) , \bar{\rho}( \ell) \right) \, , \quad
\bar{\kappa}(1)= \kappa \, ,
\nonumber \\
 \\
&&\ell \frac{\partial}{\partial \ell}  \bar{\rho} (\ell )
=\beta_\rho \left(  \bar{t}( \ell) , \bar{\kappa} (\ell ),
\bar{\sigma}( \ell) , \bar{\rho}( \ell) \right) \, , \quad
\bar{\rho}(1)= \rho \, ,
\nonumber \\
\\
&&\ell \frac{\partial}{\partial \ell}  \bar{\sigma} (\ell ) =
\beta_\sigma \left(  \bar{t}( \ell) , \bar{\kappa} (\ell ),
\bar{\sigma}( \ell) , \bar{\rho}( \ell) \right) \, , \quad
\bar{\sigma}(1)= \sigma \, .
\nonumber \\
\end{eqnarray}
\end{subequations}
In contrast to the CSE model we have to look for fixed points in a
4 dimensional parameter space. We find that the quadruple of
coupling constants  $(\bar{t}( \ell) , \bar{\kappa} (\ell ),
\bar{\rho}( \ell), \bar{\sigma}( \ell) )$ flows to the IR stable
fixed point
\begin{eqnarray}
(t^\ast , \kappa^\ast,  \rho^\ast , \sigma^\ast ) = \left
(\frac{64}{59} \sqrt{6} \, \pi \, \varepsilon , 1, 1, 0 \right)\,
.
\end{eqnarray}
This fixed point is characterized by $C_2/C_1 = \kappa^* =1$ and
$C_3/C_1= \rho^* = 1$. It turns out that the leading scaling
behavior of physical quantities depends not only on the fixed
point but also on the approach to the fixed point described by the
dangerous irrelevant variable $\sigma$.  Paths to the fixed point
decay quickly to the line described by
\begin{subequations}
\label{approach}
\begin{eqnarray}
\bar{\kappa} (\ell ) - \kappa^\ast  = (a/2 + 1/4) \, \bar{\sigma}(
\ell)
\end{eqnarray}
and
\begin{eqnarray}
\bar{\rho}( \ell) - \rho^\ast = a \, \bar{\sigma}( \ell)
\end{eqnarray}
\end{subequations}
for small $\ell$, where $a$ is an arbitrary constant. In addition
to the stable fixed point, there is the unstable Gaussian fixed
point $t^\ast =0$ and there are two unstable fixed lines which can
be parameterized as $(t(\sigma), \kappa (\sigma), \rho (\sigma),
\sigma)$ with
\begin{subequations}
\begin{eqnarray}
t (\sigma) &=& \frac{64 \, \pi \, \varepsilon \, \sqrt{2 -2
\sqrt{2\sigma} + 3 \sigma}}{3136 - 3320 \sigma + 1521\sigma^2}
\nonumber \\
&\times& \left\{ 56 \sqrt{2} + 124 \sqrt{\sigma} + 39 \sqrt{2}
\sigma \right\} \, ,
\\
\kappa (\sigma) &=& 1 - \sqrt{\sigma /2} \, ,
\\
\rho (\sigma) &=& 1 - \sqrt{2\sigma} - \sigma/2
\end{eqnarray}
\end{subequations}
and
\begin{subequations}
\begin{eqnarray}
t (\sigma) &=& \frac{64 \, \pi \, \varepsilon \, \sqrt{2 +2
\sqrt{2\sigma} + 3 \sigma}}{3136 - 3320 \sigma + 1521\sigma^2}
\nonumber \\
&\times& \left\{ 56 \sqrt{2} - 124 \sqrt{\sigma} + 39 \sqrt{2}
\sigma \right\} \, ,
\\
\kappa (\sigma) &=& 1 + \sqrt{\sigma /2} \, ,
\\
\rho (\sigma) &=& 1 + \sqrt{2\sigma} - \sigma/2 \, .
\end{eqnarray}
\end{subequations}

With help of the characteristics a formal solution to the RGE is
easily obtained,
\begin{eqnarray}
\label{solRGE}
&&\Gamma^{(M,N)} \left(  \left\{  \brm{q}_\perp, q_d \right\} ; t,
\kappa , \rho , \sigma, \mu  \right) = \bar{Z}( \ell)^{ - (M+1)/2}
 \\
&& \times \, \Gamma^{(M,N)} \Big(  \Big\{ \brm{q}_\perp,
\bar{Z}(\ell)^{-1/2} q_d \Big\} ; \bar{t} (\ell), \bar{\kappa}
(\ell) , \bar{\rho} (\ell) , \bar{\sigma} (\ell), \mu \ell \Big)
\, . \nonumber
\end{eqnarray}
Dimensional analysis gives
\begin{eqnarray}
\label{da}
&&\Gamma^{(M,N)} \left(  \left\{  \brm{q}_\perp, q_d \right\} ; t,
\kappa , \rho , \sigma, \mu  \right) =  \mu^{-(d+1) + dM + (d+1)N}
\nonumber \\
&& \times \, \Gamma^{(M,N)} \Big(  \Big\{
\frac{\brm{q}_\perp}{\mu}, \frac{q_d}{\mu^{2} } \Big\} ; t, \kappa
, \rho , \sigma, 1 \Big) \, .
\end{eqnarray}
Equation~(\ref{solRGE}) in conjunction with (\ref{da}) finally
reveals the scaling behavior of the vertex functions:
\begin{eqnarray}
\label{skalenverhalten}
&&\Gamma^{(M,N)} \left(  \left\{  \brm{q}_\perp, q_d \right\} ; t,
\kappa , \rho , \sigma, \mu  \right)
\nonumber \\
&&= \, (\mu \ell)^{-(d+1) + dM + (d+1)N}  \bar{Z}( \ell)^{ -
(M+1)/2}
 \\
&& \times \, \Gamma^{(M,N)} \Big(  \Big\{ \frac{\brm{q}_\perp}{\mu
\ell}, \frac{q_d}{(\mu \ell)^{2} \bar{Z}(\ell)^{1/2} } \Big\} ;
\nonumber \\
&& \hspace{4cm} \bar{t} (\ell), \bar{\kappa} (\ell) , \bar{\rho}
(\ell) , \bar{\sigma} (\ell), \mu \ell \Big) \, . \nonumber
\end{eqnarray}

\subsubsection{Scaling 2: physical quantities}
Now we switch back from the convenient scaled variables that we
used in our calculations to the original variables in which we
formulated our Hamiltonian~(\ref{HamilNE}). Once more, we mark
rescaled variables as well as vertex functions of the rescaled
variables with hats. Recalling our manipulations leading to
Eq.~(\ref{fieldHamilNE}) we write
\begin{subequations}
\begin{eqnarray}
&&\hat{T} = \sqrt{\frac{C_4}{K^3}} T  \, , \ \hat{t} =
\sqrt{\frac{C_4}{K^3}} t \, ,
 \\
&& \hat{q}_d = \frac{C_4}{K} q_d \, , \ \hat{u}_a =
\frac{C_4}{K}u_a \, , \  \hat{u}_d = \sqrt{\frac{C_4}{K}} u_d \, .
\quad
\end{eqnarray}
\end{subequations}
The relation between the physical vertex functions and the vertex
functions in the scaled variables is given by
\begin{eqnarray}
\label{relNE}
&& \Gamma^{(M,N)} \left( \left\{ {\rm{\bf q}}_\perp ,q_d \right\}
;  t, \kappa, \sigma , \rho , \mu \right)
\nonumber \\
&& = \, K^{M + 3N/2 -7/2} C_4^{3/2-N/2}
\nonumber \\
&& \times \, \hat{\Gamma}^{(M,N)} \left( \left\{ {\rm{\bf
q}}_\perp ,\hat{q}_d \right\} ;   \hat{t}, \kappa ,\sigma , \rho ,
\mu \right) \, .
\end{eqnarray}

Equation~(\ref{relNE}) in conjunction with
Eq.~(\ref{skalenverhalten}) gives
\begin{eqnarray}
\label{physResNE}
&&\Gamma^{(M,N)} \left( \left\{ {\rm{\bf q}}_\perp ,q_d \right\} ;
t, \kappa, \sigma , \rho , \mu \right)
 \\
&&=  \frac{1}{T} K^{M + 3N/2 -2} C_4^{1-N/2}  L_\perp^{2(d-1) -dM
-(d+1)N}
\nonumber \\
&&\times \,  \ell^{- (d+1) + dM + (d+1) N}  \left[ \bar{t}(\ell)/t
\right]^{-1} \bar{Z}(\ell)^{- (M +1)/2}
 \nonumber \\
&& \times \, \hat{\Phi}^{(M,N)} \bigg( \bigg\{ \frac{ L_\perp
{\rm{\bf q}}_\perp}{\ell} ,\frac{ L_d
q_d}{\bar{Z}(\ell)^{1/2}\ell^{2}} \bigg\} ;
\nonumber \\
&& \hspace{3cm} \sqrt{\frac{C_4}{K^3}} \bar{t} (\ell),
\bar{\kappa} (\ell), \bar{\sigma} (\ell) , \bar{\rho} (\ell) ,  1
\bigg) \, , \nonumber
\end{eqnarray}
where we introduced the susceptibilities
\begin{eqnarray}
\hat{\Phi}^{(M,N)} = \hat{t} \, \, \hat{\Gamma}^{(M,N)}
\end{eqnarray}
and where we switched from $\mu$ to the length scales
\begin{subequations}
\begin{eqnarray}
L_\perp &=& \mu^{-1}  \, ,
\\
\label{defLDNE}
L_d &=& \sqrt{\frac{C_4}{K}} \, \mu^{-2} = \sqrt{\frac{C_4}{K}} \,
L_\perp^2 \, .
\end{eqnarray}
\end{subequations}

The behavior of the elastic moduli is now easily extracted from
the 2-point vertex functions. Upon specializing
Eq.~(\ref{physResNE}) to $N=2$ and $M=0$ etc.\ we obtain
\begin{subequations}
\label{constellNE}
\begin{eqnarray}
\label{resKNE}
K (\ell ) &=& K\,  \ell^{-\varepsilon} \bar{Z}(\ell)^{-1/2} \left[
\bar{t}(\ell)/t \right]^{-1}   ,
\\
C_1 (\ell ) &=& C_1 \, \ell^{-\varepsilon} \bar{Z}(\ell)^{-3/2}
\left[ \bar{t}(\ell)/t \right]^{-1} \left[ \bar{
\sigma}(\ell)/\sigma \right]^{-1}  , \qquad
\\
C_4 (\ell ) &=& C_4\,  \ell^{-\varepsilon} \bar{Z}(\ell)^{-3/2}
\left[ \bar{t}(\ell)/t \right]^{-1} ,
\\
C_2 (\ell) &\sim& C_3 (\ell) \sim C_1 (\ell) \, .
\end{eqnarray}
\end{subequations}
For $C_1$, $C_2$ and $C_3$ we can draw our final conclusions
without solving the characteristics. Taking the derivative with
respect to the flow parameter we find, for example,
\begin{eqnarray}
\ell \frac{\partial}{\partial \ell} \, C_1 (\ell ) = - \varepsilon
- \frac{3}{2} \, \gamma - \frac{\beta_t}{t}  -
\frac{\beta_\sigma}{\sigma}  = - \gamma_\omega \, .
\end{eqnarray}
Upon expressing $\gamma_\omega$ in terms of $\sigma$, $\rho$, and
$\kappa$ and by taking into account Eqs.~(\ref{approach}) we see
that
\begin{eqnarray}
\gamma_\omega \sim \sqrt{\sigma}
\end{eqnarray}
in the vicinity of the IR-stable fixed point. Hence,
$\gamma_\omega$ vanishes in the scaling limit. Consequently,
$C_1$, $C_2$ and $C_3$ are normal.

\subsubsection{Behavior of the elastic constants for $d<3$}
For $d<3$ we can assign physical content to the length scale
$L_\perp$ by exploiting the definition of the rescaled
dimensionless temperature $\hat{t}$. This provides us with
\begin{eqnarray}
\label{LperpNE}
L_\perp =  \left( \frac{K^{3/2} \, t}{C_4^{1/2} \, T}
\right)^{1/\varepsilon}
\end{eqnarray}
as the transversal length scale associated with the onset of
scaling behavior. The corresponding longitudinal length scale can
be inferred from Eq.~(\ref{defLDNE}).

As solutions to the characteristics we obtain the power laws
\begin{subequations}
\label{powerLawsNE}
\begin{eqnarray}
\label{ergebnisZ}
\bar{Z} (\ell) &\sim& \ell^{\gamma^\ast}\, ,
\\
\bar{\sigma} (\ell) &\sim& \sigma \ell^{w}\, ,
\end{eqnarray}
where
\begin{eqnarray}
\gamma^\ast = \gamma (t^\ast, \kappa^\ast, \rho^\ast,
\sigma^\ast)= - 42 \varepsilon /59
\end{eqnarray}
and
\begin{eqnarray}
w = 4\varepsilon /59
\end{eqnarray}
\end{subequations}
is a Wegner exponent corresponding to the smallest eigenvalue of
the Hessian at the stable fixed point.

Equation~(\ref{physResNE}) in conjunction with the power-law
behavior of $\bar{Z} (\ell)$ and $\bar{\sigma} (\ell)/\sigma$
along with Eqs.~(\ref{LperpNE}) and (\ref{defLDNE}) for the
crossover-lengths $L_\perp$ and $L_d$ provide us with a complete
scaling picture of our vertex functions in $d<3$. This picture is
summarized in Eqs.~(\ref{NEscaleForms}) with
\begin{subequations}
\begin{eqnarray}
&&\eta_K =  \varepsilon + \gamma^\ast/2 = 38\varepsilon/ 59\, ,
\\
&&\eta_C = \varepsilon + 3 \gamma^\ast/2 = 4\varepsilon /59= w \,
,
\\
&&\phi = 2 + \gamma^\ast/2 = 2 - 21\varepsilon/59 \, .
\end{eqnarray}
\end{subequations}
Finally, the power-laws~(\ref{powerLawsNE}) along with
Eqs.~(\ref{constellNE}) result in the scaling forms for the
elastic constants summarized in Eqs.~(\ref{finResC4NE}) and
(\ref{finResKNE}). Note that K diverges at long length scales.
$C_4$, on the other hand, vanishes in this regime.

\subsubsection{Logarithmic behavior in $d=3$}
In 3 dimensions the solutions to the characteristics that we need
to determine the behavior of the elastic moduli are given at
leading order by
\begin{eqnarray}
\bar{t}( \ell) /t = \left[1 - \frac{7 \sqrt{6} \, t}{64  \pi} \ln
(\ell ) \right]^{-1}
\end{eqnarray}
and $\bar{Z} (\ell) \sim [\bar{t}( \ell) /t]^{- \frac{42}{59} }$.
Inserting these logarithmic solutions into Eqs.~(\ref{constellNE})
we obtain
\begin{subequations}
\label{logNE}
\begin{eqnarray}
&&K (\brm{q}_\perp ) \sim K \, \bigg[ 1 - \frac{7 \sqrt{6} \,
t}{64 \pi} \ln \left( L_\perp | \brm{q}_\perp| \right)
\bigg]^{38/59}  ,
\\
&&C_4 (\brm{q}_\perp ) \sim C_4 \, \bigg[ 1 - \frac{7 \sqrt{6} \,
t}{64  \pi} \ln \left( L_\perp | \brm{q}_\perp| \right)
\bigg]^{-4/59}  . \qquad
\end{eqnarray}
\end{subequations}
Using $\bar{\sigma} (\ell) \sim [\bar{t}( \ell) /t]^{
\frac{4}{59}}$ one can check explicitly that $C_1$, $C_2$, and
$C_3$ are normal. $K$ and $C_4$ diverge and vanish, respectively,
for $L_\perp | \brm{q}_\perp| \to 0$. Note that Eqs.~(\ref{logNE})
imply the existence of the crossover-length scale
\begin{eqnarray}
\xi_\perp = L_\perp \exp \left( \frac{64 \pi}{7 \sqrt{6} \, t}
\right) = L_\perp \exp \left( \frac{64 \, \pi \, \sqrt{K^3}}{7
\sqrt{6C_4} \, T }  \right) \, . \quad
\end{eqnarray}

\subsubsection{Poisson ratios}
Since the momentum dependence of the anomalous elastic moduli is
logarithmic in 3 dimensions it will be difficult to observe the
anomalous elasticity of NEs upon measuring these moduli directly.
However, our analysis reveals the existence of several Poisson
ratios of the elastic moduli that should be conveniently
accessible by experiments.

The IR stable fixed point values of $\kappa$, $\rho$ and $\sigma$
directly imply the Poisson ratios $C_2/C_1=1$, $C_3/C_1 = 1$ and
$C_4/C_1 = 0$. Remarkably, not only the stable fixed point but
also the approach to it contains information on relations between
the elastic moduli. From Eqs.~(\ref{approach}) it follows that
$2\kappa - \rho - 1 = \sigma/2$ for small flow parameter $\ell$.
Switching back to the original elastic constants we find the
Poisson ratio
\begin{eqnarray}
\label{apprRatio}
\frac{2\, C_2 - C_3 -C_1}{C_4} = \frac{1}{2} \, .
\end{eqnarray}
Note that Eq.~(\ref{apprRatio}) implies the Poisson ratio
$\mu/\mu_L =8$ found by Xing and
Radzihovsky~\cite{stenull_lubensky_epl2003}, where $\mu_L$ and
$\mu$ are longitudinal and transverse shear moduli,
respectively~\cite{footnoteShearModuli}.

\subsubsection{Semi-soft elasticity}
So far we considered soft elastic NEs synthesized by crosslinking
in the isotropic phase. If a NE is crosslinked in the nematic
phase, a memory of the anisotropy at the time of crosslinking is
locked in, or in other words, the rotational symmetry in the
reference space is broken. The simplest way of modeling  this
symmetry breaking is by introducing an aligning external field, or
more precisely, an aligning external stress. In this spirit we
supplement the elastic energy~(\ref{originalNEenergy}) with the
term stated in Eq.~(\ref{breakingTerm}). Next, we switch to the
relative strain~(\ref{relStrain}) and exploit a Ward identity that
corresponds at 0-loop order to
\begin{eqnarray}
\label{relHB5}
h = 2\, s\, b_5  \, .
\end{eqnarray}
By virtue of this Ward identity,  which is derived
Appendix~\ref{app:ward2}, we can study the effects of the external
aligning stress via studying the RG behavior of $b_5$ featured in
Eq.~(\ref{expansionHst1}).

Before embarking on a perturbation calculation we recall the
rescalings that led from Eq.~(\ref{expansionHst1}) to our final NE
Hamiltonian~(\ref{fieldHamilNE}). By applying these rescalings to
Eq.~(\ref{expansionHst1}) with now $b_5 = h/s$ instead of $b_5
=0$, by incorporating bending and by dropping irrelevant terms we
derive the model Hamiltonian \begin{eqnarray}
\frac{\mathcal{H}_e}{T} = \frac{\mathcal{H}}{T}+ \frac{e}{2\, T}
\int d^{d_\perp} x_\perp \int dx_d \, \partial_a u_d  \partial_a
u_d  \, ,
\end{eqnarray}
with $\mathcal{H}/T$ given by Eq.~(\ref{fieldHamilNE}). The
coupling constant $e$ is defined as
\begin{eqnarray}
e = \frac{h}{4\, K\, (1-2s)}  \, .
\end{eqnarray}
Being interested in semi-soft behavior we assume that $h$ is
small. Hence, we treat $e$ as a relevant perturbation.

Our diagrammatic calculation proceeds in much the same way as the
one outlined in Sec.~\ref{pertC5}. The only difference is that we
here have to replace the vertices~(\ref{CSEvertices}) by the NE
vertices~(\ref{NEvertices}). Expansion of the diagrams to first
order in $e$ leads to an  $\varepsilon$ pole in
$\Gamma^{(0,2)}_{dd}$ that is proportional to $e \,
\brm{q}_\perp^2$. We remove the divergence by letting
\begin{eqnarray}
e \to \mathaccent"7017{e} &=&  Z_T^{-1} Z_e e
\end{eqnarray}
with
\begin{eqnarray}
Z_e  &=& 1 -  t \, \frac{\sqrt{2+f} \, (3+f-2g+\omega)}{ 16 \, \pi
\, \varepsilon \, \sqrt{-g^2 + (2+f)\omega} }  \, .
\end{eqnarray}
The appropriate RGE here reads
\begin{eqnarray}
\label{RGEeNE}
&&\left[ D_{\mu} + e \, \psi \, \partial_e -  \left( M +
\frac{N}{2} \right) \gamma \right]
\nonumber \\
&& \times \, \Gamma^{(M,N)} \left( \left\{ {\rm{\bf q}}_\perp ,q_d
\right\} ;   t, \kappa, \rho ,\sigma ,  e, \mu \right) = 0
\end{eqnarray}
where it is understood the vertex functions are expanded to first
order in $e$. The RG differential operator is $D_\mu$ is stated in
Eq.~(\ref{diffOpNE}) and
\begin{eqnarray}
\psi = \mu \partial_\mu \ln e |_0 = \gamma_T - \gamma_e\, .
\end{eqnarray}
 In $d<3$ the characteristic for $e$,
\begin{eqnarray}
\ell \frac{\partial}{\partial \ell} \ln \bar{e} (\ell )&=& \psi
\left(  \bar{t}( \ell) , \bar{\kappa}( \ell) , \bar{\rho}( \ell),
\bar{\sigma}( \ell) \right) \, , \ \bar{e}(1)= e \, , \qquad
\end{eqnarray}
has the fixed point solution
\begin{eqnarray}
\bar{e} (\ell) \sim e \, \ell^{\psi^\ast} \, ,
\end{eqnarray}
with $\psi^\ast = \psi (t^\ast, \kappa^\ast, \rho^\ast,
\sigma^\ast) = 18 \varepsilon/59$. Supplementing the solution to
the RGE~(\ref{RGEeNE}) with a dimensional analysis and switching
to original variables yields for the physical vertex function
$\Gamma^{(0,2)}_{dd}$ the scaling form
\begin{eqnarray}
\label{physSoleNE}
&&\Gamma^{(0,2)}_{dd} \left(  {\rm{\bf q}}_\perp ,q_d ;  e \right)
= \frac{K}{T} \, L_\perp^{-4} \ell^{4- \eta_K}
\\
&&\times \, \hat{\Phi}^{(0,2)}_{dd} \left(  \frac{L_\perp{\rm{\bf
q}}_\perp}{ \ell} , \frac{L_d q_d}{ \ell^\phi} ;   \frac{L_\perp^2
e}{\ell^{1/\nu_h}} \right) \, , \nonumber
\end{eqnarray}
where we have simplified the notation by dropping several
arguments and where
\begin{eqnarray}
\nu_h= (2 -\psi^\ast)^{-1}= 1/2 +9\, \varepsilon/108  \, .
\end{eqnarray}
We emphasize the different roles played by $C_5$ and $h$ in the
CSE model and in NEs, respectively. $C_5$ corresponds to a
temperature whereas $h$ corresponds to an external magnetic field.
Nevertheless, it is useful  to introduce a length scale
\begin{eqnarray}
\xi_h = L_\perp \, (e L_\perp^2)^{-\nu_h} \sim h^{-\nu_h}
\end{eqnarray}
at this point. In the limit $\xi_h |{\rm{\bf q}}_\perp| \ll 1$ we
obtain upon choosing $\ell =(\xi_h/L_\perp)^{-\nu_h}$ that
\begin{eqnarray}
&&\Gamma^{(0,2)}_{dd} \left(  {\rm{\bf q}}_\perp ,q_d ;  e \right)
\\
&& \sim \frac{1}{T}  \left\{
\begin{array}{lcl}
K L_\perp^{-\eta_K} \xi_h^{-(2-\eta_K)}  \brm{q}_\perp^2 \sim
h^{\gamma_h} \brm{q}_\perp^2   & \quad \mbox{if}  & \quad  q_d =0
\\
C_1 q_d^2  &\quad \mbox{if} &\quad \brm{q}_\perp = \brm{0}
\end{array}
\right. \nonumber
\end{eqnarray}
where
\begin{eqnarray}
\gamma_h = \nu_h\, (2-\eta_K) = 1 -10\, \varepsilon/59 \, .
\end{eqnarray}
In the limit $\xi_h |{\rm{\bf q}}_\perp| \gg 1$, corresponding to
the limit $h \to 0$, we retrieve essentially the scaling form that
we had for $h=0$. Now, however, there is an additional correction
term that vanishes as $(\xi_h  |{\rm{\bf q}}_\perp|)^{-1/\nu_h}$.
Physically, this  correction term modifies the behavior from soft
to semi-soft. For a review of the complete scaling results, see
Eqs.~(\ref{NEscaleForms}), (\ref{finResC4NE}) and
(\ref{finResKNE}).

\section{Concluding remarks}
\label{summary}
In this paper, we have explored the anomalous elasticity of two
models for soft uniaxial elastomers, both of which are
characterized by the vanishing of the shear modulus $C_5$ for
shears in planes containing the anisotropy axis.  The first model,
which we refer to as a critically soft elastomer (CSE), describes
a uniaxial system at a simplified critical point separating a true
uniaxial elastic phase characterized by five elastic moduli and a
lower symmetry phase produced by shearing the uniaxial solid.  The
second model describes nematic elastomers (NEs) formed via
spontaneous symmetry breaking from an anisotropic state.  In the
CSE model, a bending modulus diverges, and three of the four
elastic moduli vanish as power-laws in wavenumber at long
wavelength below spatial dimension $d=3$ and logarithmically at
$d=3$.  In the NE model, the relevant bending modulus diverges,
but only one elastic constant exhibits singular behavior. In both
models, we studied the effect of turning on couplings that take
them from soft to true uniaxial elastomers.  These fields act like
temperature or external fields at a thermal critical point,
respectively, and introduce coherence lengths that diverge as a
power law as the fully soft state is approached.

The logarithmic corrections that we predict will be very difficult
to measure. Our universal Poisson ratios, on the other hand,
should be observable in experiments on three-dimensional soft
elastomers. The anomalous elasticity will be more important in 2
than in 3 dimensions. It would be interesting to find a
two-dimensional realization of nematic elastomers, say in a
crosslinked membrane confined to a substrate which inhibits height
fluctuations~\cite{xing&Co_membranes}. Another two-dimensional
system belonging the NE universality class would be a tethered
nematic membrane that orients in a plane perpendicular to an
external aligning field.

Our analysis exemplifies the power and the beauty of the
renormalization group. The renormalization group handles the
rotational invariances of nematic elastomers in two distinct
spaces quite naturally. Though the constraints imposed by these
invariances lead to almost baroquely complicated formulae at
intermediate stages of the analysis, our final results have a
surprisingly clear and simple stature.

\begin{acknowledgments}
We gratefully acknowledge financial support by the Emmy
Noether-Programm of the Deutsche Forschungsgemeinschaft (OS) and
the National Science Foundation (TCL, grant DMR 00-96531). We
thank H. K. Janssen and J. Toner for fruitful discussions. We are
particularly grateful for discussions with X. Xing, and L.
Radzihovsky, who have also studied this problem.

\end{acknowledgments}

\appendix
\section{Ward identities for critically soft elastomers}
\label{app:ward1}
By virtue of the mixing invariance discussed a the end of
Sec.~\ref{iseModel} there exist several Ward identities for CSEs.
In this Appendix we derive these identities.

To facilitate our derivation, we introduce external fields via
\begin{eqnarray}
\label{fieldstep}
\mathcal{H} \to \mathcal{H}_h = \mathcal{H} -  \int d^d x \,
\Big\{  h_a u_a + h_d u_d \Big\} \, ,
\end{eqnarray}
where $ \int d^d x = \int d^{d_\perp} x_\perp \int dx_d$. As
usual, the external fields allow us to exploit the free energy
\begin{eqnarray}
F[ h_a , h_d ] = - T \ln Z [h_a , h_d] \, ,
\end{eqnarray}
where the partition function is given by
\begin{eqnarray}
Z [h_a , h_d] = \int \mathcal{D} u_a \mathcal{D} u_d \exp \left( -
\mathcal{H}_h /T\right) \, ,
\end{eqnarray}
as a generating function for the order parameters
\begin{eqnarray}
m_a (\brm{x}) = \langle u_a (\brm{x})  \rangle = \frac{\delta
F}{\delta h_a (\brm{x})} \, ,
\\
m_d (\brm{x}) = \langle u_d (\brm{x})  \rangle = \frac{\delta
F}{\delta h_d (\brm{x})} \, .
\end{eqnarray}
As a consequence of the mixing invariance, the free energy obeys
the relation
\begin{eqnarray}
\label{relF}
F[ h_a , h_d ] =  \theta_a  \int d^d x \, x_a h_d + F[ h_a ,
h_d^\prime ]  \, ,
\end{eqnarray}
where $h_d^\prime = h_d + \theta_a h_a$. Because the right hand
side of Eq.~(\ref{relF}) is independent of the $\theta_a$, we
obtain
\begin{eqnarray}
\label{defL}
&&L_a = \frac{\partial F}{\partial \theta_a} =  \int d^d x \, x_a
h_d (\brm{x}) + \int d^d x \,  m_d (\brm{x}) h_a (\brm{x})
\nonumber \\
&& \, = 0 \, .
\end{eqnarray}

Now we adopt the usual strategy and take various derivatives. We
start with
\begin{eqnarray}
&&\frac{ \delta^2 L_a}{\delta m_d (\brm{y}) \, \delta m_b
(\brm{z})} =   \int d^d x \, x_a \,  \frac{ \delta^2  h_d
(\brm{x})}{\delta m_d (\brm{y}) \, \delta m_b (\brm{z})}
\nonumber \\
&& \, +  \frac{ \delta  h_a (\brm{y})}{ \delta m_b (\brm{z})} +
\int d^d x \,  m_d (\brm{x}) \,  \frac{ \delta^2  h_a
(\brm{x})}{\delta m_d (\brm{y}) \, \delta m_b (\brm{z})} = 0 \, .
\end{eqnarray}
In equilibrium, the order parameters vanish for vanishing external
fields, $h_a = h_d = 0$. Hence, we get the Ward identity
\begin{eqnarray}
\label{WT1a}
- \int d^d x \, x_a \, \Gamma^{(1,2)}_{ddb} \left( \brm{x},
\brm{y}, \brm{z} \right) = \Gamma^{(2,0)}_{ab} \left( \brm{y},
\brm{z} \right) \, .
\end{eqnarray}
In Fourier space, identity~(\ref{WT1a}) takes the form
\begin{eqnarray}
\label{WT1b}
-i \frac{\partial}{\partial p^{(1)}_a} \, \Gamma^{(1,2)}_{ddb}
\big( \brm{p}^{(1)}, \brm{p}^{(2)}, \brm{p}^{(3)} \big)
\Big|_{\brm{p}^{(1)} = \brm{0}} = \Gamma^{(2,0)}_{ab} \big(
\brm{p}^{(2)}, \brm{p}^{(3)} \big)
\nonumber \\
\end{eqnarray}
with $\brm{p}^{(2)} +  \brm{p}^{(3)} = \brm{0}$. Another Ward
identity can be found by taking the second functional derivative
of $L_a$ with respect to $m_d (\brm{y})$ and $m_d (\brm{z})$. This
leads in Fourier space to
\begin{eqnarray}
\label{WT2}
&& -i \frac{\partial}{\partial p^{(1)}_a} \, \Gamma^{(0,3)}_{ddd}
\big( \brm{p}^{(1)}, \brm{p}^{(2)}, \brm{p}^{(3)} \big)
\Big|_{\brm{p}^{(1)} = \brm{0}}
\nonumber \\
&&=  \, \Gamma^{(1,1)}_{ad} \big(  \brm{p}^{(2)}, \brm{p}^{(3)}
\big) +  \Gamma^{(1,1)}_{ad} \big(  \brm{p}^{(3)}, \brm{p}^{(2)}
\big) \, .
\end{eqnarray}
An identity for the four point vertex function
$\Gamma^{(0,4)}_{dddd}$ follows in a similar manner upon taking a
third order functional derivative of $L_a$ with respect to, say,
$m_d (\brm{y})$, $m_d (\brm{z})$, and $m_d (\brm{w})$. The result
can be stated as
\begin{eqnarray}
\label{WT3}
&&-i \frac{\partial}{\partial p^{(1)}_a} \, \Gamma^{(0,4)}_{dddd}
\big( \brm{p}^{(1)}, \brm{p}^{(2)}, \brm{p}^{(3)},  \brm{p}^{(4)}
\big) \Big|_{\brm{p}^{(1)} = \brm{0}}
\nonumber \\
&& = \,  \Gamma^{(1,2)}_{add} \big( \brm{p}^{(2)}, \brm{p}^{(3)},
\brm{p}^{(4)} \big) + \Gamma^{(1,2)}_{add} \big( \brm{p}^{(4)},
\brm{p}^{(2)}, \brm{p}^{(3)} \big)
\nonumber \\
&& +\,  \Gamma^{(1,2)}_{add} \big( \brm{p}^{(3)}, \brm{p}^{(4)},
\brm{p}^{(2)} \big) \, .
\end{eqnarray}

Of course, our Ward identities should hold to arbitrary order in
perturbation theory. At zero loop order, the consistency of the
Ward identities can be checked without much effort. We carried out
this check and found our identities confirmed.

\section{Ward identities for nematic elastomers I}
\label{app:ward2}
Here we derive the Ward identities that follow from the reference
space rotation invariance of the NE
Hamiltonian~(\ref{originalNEenergy}). One of these Ward identities
can be viewed as the origin of the soft elasticity characterizing
NEs crosslinked in the isotropic phase. In addition, we consider
semi-soft NEs where the reference space rotation invariance is
broken by an aligning external stress.

\subsection{The soft case}
\label{app:ward2Soft}
To make our arguments more intuitive we work directly in $d=3$
dimensions. Without loss of generality we assume rotations about
the $x$ axis which we parameterize by a rotation angle $\theta$.
For small $\theta$ these rotations are described by the orthogonal
matrix
\begin{eqnarray}
\tens{O}_R = \left(
\begin{array}{ccc}
1 & 0& 0
\\
0 & 1 & \theta
\\
0 & -\theta & 1
\end{array}
 \right) \, .
\end{eqnarray}
Such a rotation takes the strain $\tens{u}$ to $\tens{u}^\prime =
\tens{O}_R \, \tens{u} \, \tens{O}_R^T$, where $\tens{O}_R^T$ is
the transposed of $\tens{O}_R$. The relative strain $\tens{w} =
\tens{u} - \tens{u}_0$ is taken by the rotation to
\begin{eqnarray}
\tens{w}^\prime = \tens{O}_R \, \tens{u}_0 \, \tens{O}_R^T  -
\tens{u}_0 + \tens{O}_R \, \tens{w} \, \tens{O}_R^T \, .
\end{eqnarray}

We introduce an external symmetric stress $\tens{\sigma}$ into our
model via
\begin{eqnarray}
\mathcal{H} \to \mathcal{H}_\sigma = \mathcal{H} -  \int d^d x \,
\sigma_{ij} w_{ij} \, ,
\end{eqnarray}
where $\mathcal{H}$ is the elastic energy of
Eq.~(\ref{expansionHst1}) supplemented with the relevant bending
term. Then, the free energy
\begin{eqnarray}
F\left[  \tens{\sigma} \right] = - T \ln Z
\left[\tens{\sigma}\right] \, ,
\end{eqnarray}
with the partition function
\begin{eqnarray}
Z \left[\tens{\sigma}\right] = \int \mathcal{D} \tens{w} \exp
\left(  - \mathcal{H}_\sigma /T\right) \, ,
\end{eqnarray}
is a generating function for the tensor order parameter\
\begin{eqnarray}
m_{ij} (\brm{x}) = \langle w_{ij} (\brm{x})  \rangle =
\frac{\delta F}{\delta \sigma_{ij} (\brm{x})} \, .
\end{eqnarray}
Owing to the rotation invariance of the
Hamiltonian~(\ref{expansionHst1}) without external stress the free
energy satisfies
\begin{eqnarray}
\label{relFNE}
F\left[  \tens{\sigma} \right] =  - \theta \, s   \int d^d x  \,
[\sigma_{23} + \sigma_{32}]  +  F\left[  \tens{\sigma}^\prime
\right]  \, ,
\end{eqnarray}
with the elements of the symmetric tensor $\tens{\sigma}^\prime$
given by
\begin{subequations}
\begin{eqnarray}
\sigma^\prime_{11} &=& \sigma_{11} \, ,
\\
\sigma^\prime_{12} &=& \sigma_{12} - \theta \, \sigma_{13} \, ,
\\
\sigma^\prime_{13} &=& \sigma_{13} + \theta \, \sigma_{12} \, ,
\\
\sigma^\prime_{22} &=& \sigma_{22} - \theta \, (\sigma_{23} +
\sigma_{32}) \, ,
\\
\sigma^\prime_{23} &=& \sigma_{23} + \theta \, (\sigma_{22} -
\sigma_{33}) \, ,
\\
\sigma^\prime_{33} &=& \sigma_{33} + \theta \, (\sigma_{23} +
\sigma_{32}) \, .
\end{eqnarray}
\end{subequations}

Evidently, the left hand side of the identity~(\ref{relFNE}) is
independent of the rotation angle $\theta$. Hence, the quantity
\begin{eqnarray}
\label{defLNE}
L &=& \frac{\partial F}{\partial \theta} =  \int d^d x  \, \big\{
- s \Gamma_{23} (\brm{x})  + m_{12} \Gamma_{13}(\brm{x})
\nonumber \\
&-& m_{13} \Gamma_{12} (\brm{x})   - 2 m_{23} [\Gamma_{22}
(\brm{x}) -\Gamma_{33} (\brm{x}) ] \nonumber \\
& - & [m_{33}(\brm{x})-m_{22}(\brm{x})] \Gamma_{23}(\brm{x})
\big\}
\end{eqnarray}
vanishes identically. In this expression, we used the facts that
$m_{ij} (\brm{x}) $ and $\sigma_{ij}$ are symmetric tensors and
that $\Gamma_{ij} (\brm{x}) = 2 \delta F/\delta m_{ij} (\brm{x})$
for $i \neq j$ (we use only $i<j$) and $\Gamma_{ii} (\brm{x}) =
\delta F/\delta m_{ii} (\brm{x})$ for all $i$ (no Einstein
convention). Now we are in the position to extract the sought
after Ward identities simply by taking functional derivatives of
$L$ with respect to the order parameter. For example,
differentiating with respect to $m_{23} ( \brm{y})$ and setting
$m_{ij} = 0$, we obtain the Ward identity,
\begin{eqnarray}
\label{ex1}
& & \frac{\delta L}{\delta m_{23} (\brm{y})}=
\\
& & -s \int d^d x \, \Gamma_{2323} (\brm{x}, \brm{y}) - 2
[\Gamma_{22} (\brm{y}) - \Gamma_{33} (\brm{y})] =0 , \nonumber
\end{eqnarray}
which in Fourier space is
\begin{eqnarray}
\label{ex3}
 s  \, \Gamma_{2323} (\brm{0}, \brm{0}) +2[\Gamma_{22} (\brm{0})
- \Gamma_{33} (\brm{0}) ]= 0 \, .
\label{ward1a}
\end{eqnarray}
This Ward identity is particularly important because it is the
origin of the soft elasticity of NEs. Taking the appropriate
derivatives of the free energy in Eq.\ (\ref{expansionHst1}), it
is straightforward to show that $\Gamma_{2323}(\brm{0}, \brm{0})=2
b_5$, $\Gamma_{33} (\brm{0}, \brm{0})=a_1$, and $\Gamma_{22}
(\brm{0}, \brm{0})=a_2$ and thus that Eq.\ (\ref{ward1a}) reduces
relation~(\ref{coeffRelA}) at 0-loop order.

The derivation of the other Ward identities that generalize the
remaining relations stated in Eqs.~(\ref{coeffRels}) is a similar
exercise. We restrict ourselves to stating the final results
\begin{subequations}
\begin{eqnarray}
&-& s  \, \Gamma_{232311} (\brm{0}, -\brm{p}, \brm{p}) -2
\Gamma_{2211} (-\brm{p}, \brm{p})
\nonumber \\
&+& 2 \Gamma_{3311} (-\brm{p}, \brm{p}) = 0 \, ,
 \\
&-& s  \, \Gamma_{232333} (\brm{0}, -\brm{p}, \brm{p}) -2
\Gamma_{2233} (-\brm{p}, \brm{p})
\nonumber \\
&+& 2 \Gamma_{3333} (-\brm{p}, \brm{p}) - \Gamma_{2323} (-\brm{p},
\brm{p})= 0 \,
\\
&-& s  \, \Gamma_{231312} (\brm{0}, -\brm{p}, \brm{p}) -2
\Gamma_{1212} (-\brm{p}, \brm{p})
\nonumber \\
&+& 2 \Gamma_{1313} (-\brm{p}, \brm{p}) = 0 \, ,
\\
&-& s  \, \Gamma_{23231313} (\brm{0}, \brm{p}^{(1)},
\brm{p}^{(2)}, \brm{p}^{(3)}) -2 \Gamma_{221313} ( \brm{p}^{(1)},
\brm{p}^{(2)}, \brm{p}^{(3)})
\nonumber \\
&+& 2 \Gamma_{331313} ( \brm{p}^{(1)}, \brm{p}^{(2)},
\brm{p}^{(3)}) - 2 \Gamma_{122313} ( \brm{p}^{(2)}, \brm{p}^{(1)},
\brm{p}^{(3)})
\nonumber \\
&-& \Gamma_{122313} ( \brm{p}^{(3)}, \brm{p}^{(1)},
\brm{p}^{(2)})= 0 \, ,
\end{eqnarray}
\end{subequations}

\subsection{The semi-soft case}
Now we consider semi-soft NEs modeled by the elastic
energy~(\ref{originalNEenergy}) supplemented with the explicit
uniaxial term~(\ref{breakingTerm}). The aligning stress $h$ breaks
the reference space rotation symmetry in much the same way as the
stress $\underline{\underline{\sigma}}$ utilized in
Appendix~\ref{app:ward2Soft}. Thus, we basically just have to
repeat the steps carried out in Appendix~\ref{app:ward2Soft} with
\begin{eqnarray}
\mathcal{H}_\sigma \to \mathcal{H}_{h,\sigma} &=& \mathcal{H}_h -
\int d^d x \, \sigma_{ij} w_{ij}
\nonumber \\
&=& \mathcal{H} -  \int d^d x \, \sigma_{h,ij} w_{ij} \, ,
\end{eqnarray}
where
\begin{eqnarray}
\sigma_{h,ij} = \sigma_{ij} + \frac{h}{d} \left( \delta_{i1}
\delta_{j1} +  \delta_{i2}  \delta_{j2} \right) -
\frac{(d-1)\,h}{d} \delta_{i3}  \delta_{j3} \, .
\nonumber \\
\end{eqnarray}
This analysis leads in Fourier space to the Ward identity
\begin{eqnarray}
\label{semiSoftWard}
 s  \, \Gamma_{2323} (\brm{0}, \brm{0}) -h  = 0
\end{eqnarray}
provided that the equations of state $\Gamma_{22} (\brm{0}) = 0$
and $\Gamma_{33} (\brm{0}) = 0$ are satisfied. At 0-loop level the
identity~(\ref{semiSoftWard}) reduced to the relation between $h$
and $b_5$ stated in Eq.~(\ref{relHB5}).

\section{Ward identities for nematic elastomers II}
\label{app:ward3}
The mixing invariance of NEs leads to Ward identities analogous to
those for CSEs derived in Appendix~\ref{app:ward1}. Since the
derivation of the 2 sets of identities is similar and
Appendix~\ref{app:ward1} is fairly detailed we restrict ourselves
here to mention differences in the derivations and to list
results.

Introducing external fields via the step~\ref{fieldstep} we find
that the free energy of NEs satisfies the identity
\begin{eqnarray}
\label{relFNE2}
&&F[ h_a (\brm{x}), h_d (\brm{x})] =  -\theta_a  \int d^d x \, x_a
h_d (\brm{x})
 \\
&&+\,  F[ h_a (x_c + \theta_c x_d, x_d), h_d (x_c + \theta_c x_d,
x_d) + \theta_a h_a (\brm{x}) ]  \, . \nonumber
\end{eqnarray}
Thus, the quantity $L_a = \partial F/\partial \theta_a$ obeys the
equation
\begin{eqnarray}
\label{defLNE2}
&&L_a=  - \int d^d x \, x_a h_d (\brm{x}) + \int d^d x \,  m_d
(\brm{x}) h_a (\brm{x})
\nonumber \\
&&+\,  \int d^d x \,  m_c (\brm{x}) \frac{\partial h_c
(\brm{x})}{\partial x_a} \, x_d + \int d^d x \,  m_d (\brm{x})
\frac{\partial h_d (\brm{x})}{\partial x_a} \, x_d
\nonumber \\
&& \, = 0 \, .
\end{eqnarray}
By taking the appropriate functional derivatives with respect to
the order parameter, we obtain the Ward identities
\begin{subequations}
\begin{eqnarray}
&& i \frac{\partial}{\partial p^{(1)}_a} \, \Gamma^{(1,2)}_{ddb}
\big( \brm{p}^{(1)}, \brm{p}^{(2)}, \brm{p}^{(3)} \big)
\Big|_{\brm{p}^{(1)} = \brm{0}} = \Gamma^{(2,0)}_{ab} \big(
\brm{p}^{(2)}, \brm{p}^{(3)} \big)
\nonumber \\
&&-\,  \frac{\partial}{\partial \, p_d^{(2)}} \,
\Gamma^{(1,1)}_{bd} \big(  \brm{p}^{(2)}, \brm{p}^{(3)} \big) \,
p_a^{(2)}
\nonumber \\
&&-\, \frac{\partial}{\partial \, p_d^{(3)}} \,
\Gamma^{(1,1)}_{bd} \big(  \brm{p}^{(2)}, \brm{p}^{(3)} \big) \,
p_a^{(3)} \, ,
 \\
 && i \frac{\partial}{\partial p^{(1)}_a} \, \Gamma^{(0,3)}_{ddd} \big( \brm{p}^{(1)}, \brm{p}^{(2)}, \brm{p}^{(3)} \big) \Big|_{\brm{p}^{(1)} = \brm{0}}
 \nonumber \\
 &&= \, \Gamma^{(1,1)}_{ad} \big(  \brm{p}^{(2)}, \brm{p}^{(3)} \big) + \Gamma^{(1,1)}_{ad} \big(  \brm{p}^{(3)}, \brm{p}^{(2)} \big)
\nonumber \\
&&-\,  \frac{\partial}{\partial \, p_d^{(2)}} \,
\Gamma^{(0,2)}_{dd} \big(  \brm{p}^{(2)}, \brm{p}^{(3)} \big) \,
p_a^{(2)}
\nonumber \\
&&-\, \frac{\partial}{\partial \, p_d^{(3)}} \,
\Gamma^{(0,2)}_{dd} \big(  \brm{p}^{(2)}, \brm{p}^{(3)} \big) \,
p_a^{(3)} \, ,
\\
&& i \frac{\partial}{\partial p^{(1)}_a} \, \Gamma^{(0,4)}_{dddd}
\big( \brm{p}^{(1)}, \brm{p}^{(2)}, \brm{p}^{(3)}, \brm{p}^{(4)}
\big) \Big|_{\brm{p}^{(1)} = \brm{0}}
 \nonumber \\
 &&= \, \Gamma^{(1,2)}_{add} \big(  \brm{p}^{(2)}, \brm{p}^{(3)} , \brm{p}^{(4)}\big) + \Gamma^{(1,2)}_{add} \big(  \brm{p}^{(3)}, \brm{p}^{(2)} , \brm{p}^{(4)} \big)
\nonumber \\
&&+ \, \Gamma^{(1,2)}_{add} \big(  \brm{p}^{(4)}, \brm{p}^{(2)} ,
\brm{p}^{(3)} \big)
\nonumber \\
&&-\,  \frac{\partial}{\partial \, p_d^{(2)}} \,
\Gamma^{(0,3)}_{ddd} \big(  \brm{p}^{(2)}, \brm{p}^{(3)} ,
\brm{p}^{(4)} \big) \, p_a^{(2)}
\nonumber \\
&&-\, \frac{\partial}{\partial \, p_d^{(3)}} \,
\Gamma^{(0,3)}_{ddd} \big(  \brm{p}^{(2)}, \brm{p}^{(3)} ,
\brm{p}^{(4)} \big) \, p_a^{(3)}
\nonumber \\
&&-\, \frac{\partial}{\partial \, p_d^{(4)}} \,
\Gamma^{(0,3)}_{ddd} \big(  \brm{p}^{(2)}, \brm{p}^{(3)} ,
\brm{p}^{(4)} \big) \, p_a^{(4)} \, .
\end{eqnarray}
\end{subequations}

\section{Alternative renormalization scheme for nematic elastomers}
\label{app:alternativeScheme}
Our renormalization scheme~(\ref{NEscheme}) which involves a re-parameterization of the coordinate $x_d$ follows closely the approach developed by Grinstein and Pelcovitz~\cite{grinstein_pelcovits_81_82}. Though this approach has been established for more that 2 decades now, it is not clear how the known strategies of proving renormalizability apply to it. In this appendix we briefly present an alternative renormalization scheme for NEs that does not entail a re-parameterization of $x_d$. We demonstrate that the alternative formulation leads exactly to the same results as our original approach.

Let us revisit our Hamiltonian~(\ref{HamilNE}) as a starting point. To reduce the number of scaling variables we set $T = \check{T}\, K$. This step yields
\begin{eqnarray}
\label{fieldHamilNEalternative}
\frac{\mathcal{H}}{T} &=& \frac{1}{2\, \check{T}} \int d^{d_\perp} x_\perp
\int d x_d  \, \big\{  \check{\omega}\,  v_{dd}^2 +  \left( \nabla_\perp^2
u_d \right)^2
\nonumber \\
&+& 2 \, \check{g} \, v_{dd} v_{aa} +  \check{f} \, v_{aa}^2 + 2  \, \check{h} \, v_{ab}^2
\big\}  \, ,
\end{eqnarray}
where $\check{\omega} = C_1/K$, $\check{g} = C_2/K$, $\check{f} = C_3/K$ and $\check{h} = C_4/K$. Note that $\check{h}$ has no counterpart in Eq.~(\ref{fieldHamilNE}) and that we thus have an additional parameter in comparison to Eq.~(\ref{fieldHamilNE}). Due to this additional parameter, the Hamiltonian~(\ref{fieldHamilNEalternative}) has an additional invariance, viz.\ it is invariant in from under the rescaling $x_d \to \beta^{-1/2} x_d$, $u_d \to \beta^{1/2} u_d$, $u_a \to \beta^{1} u_a$, $\check{T} \to \beta^{1/2} \check{T}$, $\check{\omega} \to \beta^{-1} \check{\omega}$, $\check{g} \to \beta^{-1} \check{g}$, $\check{f} \to \beta^{-1} \check{f}$ and $\check{h} \to \beta^{-1} \check{h}$ ($\beta$-rescaling). This scaling invariance implies that the vertex functions obey the scaling from
\begin{eqnarray}
\label{scalingFormBeta}
&&\Gamma^{(M,N)} \left(  \left\{  \brm{q}_\perp, q_d \right\} ; \check{T}, \check{\omega}, \check{g}, \check{f}, \check{h} \right) =  \beta^{(M+1)/2}
\nonumber \\
&& \times \, \Gamma^{(M,N)} \left(  \Big\{
\brm{q}_\perp, \beta^{1/2} q_d  \Big\} ; \beta^{1/2} \check{T},  \frac{\check{\omega}}{\beta}
,\frac{\check{g}}{\beta} , \frac{\check{f}}{\beta}, \frac{\check{h}}{\beta} \right) \, . \quad
\end{eqnarray} 
Equation~(\ref{scalingFormBeta}) suggest to introduce composite parameters that are invariant under the $\beta$-rescaling. Among various possibilities we choose
\begin{subequations}
\begin{eqnarray}
&&t = \mu^{-\varepsilon} \check{h}^{1/2} \check{T} = \mu^{-\varepsilon} T  \sqrt{C_4 /K^3} \, ,
\\
&&\omega = \check{\omega} /  \check{h} = C_1 /C_4 \, ,
\\
&&g = \check{g} /  \check{h} = C_2 /C_4\, ,
\\
&&f = \check{f} /  \check{h} = C_3 /C_4\, .
\end{eqnarray}
\end{subequations}
Note that these composite parameters are identical to the parameters we introduced by switching from
Eq.~(\ref{HamilNE}) to (\ref{fieldHamilNE}).

After this prelude we specify our alternative renormalization scheme,
\begin{subequations}
\label{NEschemeAlt}
\begin{eqnarray}
&&\check{T} \to \mathaccent"7017{\check{T}}  = Z_T \, \check{T}\, ,
\\
&&\check{\omega} \to \mathaccent"7017{\check{\omega}}  = Z_T^{-1} Z_\omega \, \check{\omega} \, ,
\\
&&\check{g} \to \mathaccent"7017{\check{g}}  = Z_T^{-1} Z_g \, \check{g} \, ,
\\
&&\check{f} \to \mathaccent"7017{\check{f}}  = Z_T^{-1} Z_h \, \check{f}\, ,
\\
&&\check{h} \to \mathaccent"7017{\check{h}}  = Z_T^{-1} Z_h \, \check{h} \, .
\end{eqnarray}
\end{subequations}
With this scheme 
\begin{eqnarray}
\label{fieldHamilNEalternativeRen}
\frac{\mathcal{H}}{T} &\to& \frac{1}{2\, \check{T}} \int d^{d_\perp} x_\perp
\int d x_d  \, \big\{  Z_\omega \check{\omega}\,  v_{dd}^2 +  Z_T \left( \nabla_\perp^2
u_d \right)^2
\nonumber \\
&+& 2 \, Z_g \check{g} \, v_{dd} v_{aa} + Z_f  \check{f} \, v_{aa}^2 + 2  \, Z_h \check{h} \, v_{ab}^2
\big\}  \, .
\end{eqnarray}
Next we derive a RGE for the vertex functions. Exploiting as usual the independence of the bare theory of $\mu$ and upon switching via $t$, $\omega$, $g$ and $f$ to the benign parameters $t$, $\kappa$, $\rho$ and $\sigma$ we get
\begin{eqnarray}
\label{RGEalternative}
&&\left[ \mu \partial_\mu + \zeta \check{h} \partial_{\check{h}} + \beta_t \partial_t  + \beta_\kappa \partial_\kappa + \beta_\rho \partial_\rho + \beta_\sigma \partial_\sigma  \right]
\nonumber \\
&& \times \, \Gamma^{(M,N)} \left(  \left\{  \brm{q}_\perp, q_d \right\} ; t,
\kappa , \rho , \sigma, \check{h} , \mu  \right) = 0 \, ,
\end{eqnarray}
where
\begin{eqnarray}
\label{wilsonZetaAlt}
\zeta = \mu \partial_\mu \ln \check{h} \big|_0 = \gamma_T - \gamma_h
\end{eqnarray}
and where the Wilson $\beta$- and $\gamma$-functions are defined as usual. By solving the RGE~(\ref{RGEalternative}), supplementing the solution with a dimensional analysis and by exploiting Eq.~(\ref{scalingFormBeta}) with the choice $\beta = \check{h}$ for the parameter of the $\beta$-rescaling we find
\begin{align}
\label{scalingFormAlt}
&\Gamma^{(M,N)} \left(  \left\{  \brm{q}_\perp, q_d \right\} ; t,
\kappa , \rho , \sigma, \check{h}, \mu \right) =  \bar{\check{h}} (\ell)^{(M+1)/2}
\nonumber \\
& \times \Gamma^{(M,N)} \bigg(  \bigg\{
\frac{\brm{q}_\perp}{\mu \ell}, \frac{  \bar{\check{h}} (\ell)^{1/2} q_d}{(\mu \ell)^2  }  \bigg\} ;  \bar{t} (\ell) ,
\bar{\kappa} (\ell) , \bar{\rho}(\ell) , \bar{\sigma} (\ell), 1, \mu \ell \bigg)  . \quad
\end{align}
Now we take a closer look at the RG flow of $\check{h}$. Comparing the renormalized Hamiltonians~(\ref{fieldHamilNEalternativeRen}) and (\ref{renoHamilNE}) we learn that
\begin{eqnarray}
Z_h = Z_T\, Z \, ,
\end{eqnarray}
provided the left and the right hand side are expressed in terms of the same variables, here in particular $t$, $\kappa$, $\rho$ and $\sigma$. Using Eq.~(\ref{wilsonZetaAlt}) and the definition of the Wilson $\gamma$-functions we find
\begin{eqnarray}
\label{hEquiv}
\zeta = - \gamma \, .
\end{eqnarray}
From the corresponding characteristics, we consequentially obtain
\begin{eqnarray}
\bar{\check{h}} (\ell) \sim \bar{Z} (\ell)^{-1} \, .
\end{eqnarray}
Collecting Eqs.~(\ref{scalingFormAlt}) and (\ref{hEquiv}) we learn that our original and our alternative formulation lead to equivalent scaling results for the vertex functions. Of course, all results that follow from the scaling forms for the vertex functions, in particular the anomalous behavior of the elastic moduli, are identical for both approaches.  

\section{Calculation of Feynman diagrams}
\label{app:diagramCalc}
Here we give some details on the calculation of the Feynman
diagrams listed in Figs.~\ref{diagrams1} to \ref{diagrams3}. The
first part of this Appendix contains 2 representative examples.
All the remaining diagrams may then be computed by similar means.
The second part features 2 parameter integrals. All 2-leg diagrams
can be expressed in terms of these integrals.

\subsection{Examples}
As a first example we consider the first diagram in the first row
of Fig.~\ref{diagrams1}. For the sake of the argument we refer to
this diagram as $A_{dd}$. Regarding CSEs this diagram stands for
the mathematical formula
\begin{eqnarray}
\label{A1}
A_{dd} &=& q_d^2 \, \frac{g^2}{2 \, T^2} \,  \int_{\brm{k}}\, k_a (k_a +
q_a)  \, k_b (k_b + q_b)
\nonumber \\
&\times& G_{dd} (\brm{k}) \, G_{dd} (\brm{k}+ \brm{q})\, ,
\end{eqnarray}
where $\brm{q}$ is an external momentum running trough the
diagram. $\int_{\brm{k}}$ is an abbreviation for $1/(2\pi)^d \int
d^{d_\perp} k_\perp  \, \int d k_d$. Simple power counting reveals
that the superficial degree of divergence of $A_{dd}$ is zero.
Hence, it is sufficient to evaluate the diagram at vanishing
external momentum. For $\brm{q} = \brm{0}$, $A_{dd}$ reduces to
\begin{eqnarray}
\label{A2}
A_{dd} &=& q_d^2 \, \frac{g^2}{2} \, B^2 \,  M^{(1)}_{42}\, ,
\end{eqnarray}
with $M^{(1)}_{42}$ being a specification of the parameter
integral $M^{(1)}_{ln}$ defined in Eq.~(\ref{M11}). Exploiting the
result~(\ref{M15}) we obtain
\begin{eqnarray}
\label{A3}
A_{dd} &=& q_d^2 \, g^2 \, \frac{B^{1/2}}{A^{1/2}} \,
\frac{\tau^{-\varepsilon/4}}{16 \, \pi\, \varepsilon}
\end{eqnarray} as the final result for $A_{dd}$. Regarding NEs we
simply have to replace the $g^2$ stemming from the vertices by
$(\omega - g)^2$.

As a second example we illustrate the computation of the last
diagram in the row line of Fig.~\ref{diagrams1}. Let us call this
diagram $C_{dd}$. In case of the CSE model it visualized the
formula
\begin{eqnarray}
\label{C1}
C_{dd} &=&\frac{g^2}{T^2} \,  \int_{\brm{k}}\, k_d^2 \, \left[  \brm{q}_\perp
\cdot (\brm{q}_\perp + \brm{k}_\perp) \right]^2
\nonumber \\
&\times& G_{dd} (\brm{k}) \, G_{dd} (\brm{k}+ \brm{q})\, ,
\end{eqnarray}
The superficial degree of divergence of this diagram is 2. Hence
an evaluation at $\brm{q} = \brm{0}$ is not sufficient and we
rather have to expand $C_{dd}$ in powers of the external momentum,
Using the parameter integral $M^{(2)}_{ln}$ defined in
Eq.~(\ref{M21}) this expansion can be written as
\begin{eqnarray}
\label{C2}
C_{dd} &=& \brm{q}_\perp^4 \, g^2 \,  B^2 \Big\{  M^{(2)}_{02} -
\frac{9}{2} \, B \, M^{(2)}_{43} + 2\, B^2 \, M^{(2)}_{86} \Big\}
\, .
\end{eqnarray}
All the other terms in this expansion turn out to be UV
convergent. Using Eq.~(\ref{M25}) the final result for $C_{dd}$ is
now readily found to be
\begin{eqnarray}
\label{C3}
C_{dd} &=& \brm{q}_\perp^4 \, g^2 \,  \frac{B^{3/2}}{A^{3/2}} \,
\frac{\tau^{-\varepsilon/4}}{64 \, \pi\, \varepsilon}  \, .
\end{eqnarray}
Again, one simply has to replace the $g^2$ stemming from the
vertices by $(\omega - g)^2$ if one is interested in NEs.

\subsection{Parameter integrals}
Many of the 2-leg diagrams can be expressed in terms of the
parameter integral
\begin{eqnarray}
\label{M11}
M^{(1)}_{ln} = \int_{\brm{k}} \frac{\brm{k}_\perp^l}{[B\tau +
Ak_d^2 + B \brm{k}_\perp^4]^n} \, .
\end{eqnarray}
The calculation of this integral can be simplified by rescaling
the momenta,
\begin{eqnarray}
\label{M12}
M^{(1)}_{ln} = A^{-1/2} \, B^{-(d+l-1)/4} \, \int_{\brm{k}}
\frac{\brm{k}_\perp^l}{[B\tau + k_d^2 +  \brm{k}_\perp^4]^n} \, .
\end{eqnarray}
Next, we employ Schwinger representation,
\begin{eqnarray}
\label{M13}
&& M^{(1)}_{ln} = \frac{A^{-1/2} \, B^{-(d+l-1)/4}}{(2\pi)^d \,
\Gamma (n)} \, \int_0^\infty ds \, s^{n-1} \exp (- sB\tau)
\nonumber \\
&& \times \, \int d^{d_\perp} k_\perp  \, \brm{k}_\perp^l \exp
\left(- s \brm{k}_\perp^4 \right) \int d k_d \, \exp \left(- s
k_d^2 \right) \, .
\end{eqnarray}
Now the momentum integrations are straightforward. We obtain
\begin{eqnarray}
\label{M14}
&& M^{(1)}_{ln} = \frac{A^{-1/2} \, B^{-(d+l-1)/4} \, \Gamma
(\frac{d+l-1}{4})}{2 \, (4\pi)^{d/2} \, \Gamma (n) \, \Gamma
(\frac{d-1}{2})}
\nonumber \\
&& \times \, \int_0^\infty ds \,  \exp (- sB\tau) \, s^{-1 +
\frac{4n -l-d-1}{4}} \, .
\end{eqnarray}
Carrying out the remaining integration over the Schwinger
parameter $s$ gives finally
\begin{eqnarray}
\label{M15}
&& M^{(1)}_{ln} = \frac{A^{-1/2} \, B^{-(2n-1)/2} \, \Gamma
(\frac{2+l-\varepsilon}{4})}{2 \, (4\pi)^{d/2} \, \Gamma (n) \,
\Gamma (\frac{2-\varepsilon}{2})}
\nonumber \\
&& \times \, \Gamma \left( \frac{4n -l-4 +\varepsilon}{4} \right)
\tau^{- \frac{4n -l-4 +\varepsilon}{4}} \, .
\end{eqnarray}

In addition to $M^{(1)}_{ln}$ it turns out to be convenient to
introduce a second parameter integral, viz.\
\begin{eqnarray}
\label{M21}
M^{(2)}_{ln} = \int_{\brm{k}} \frac{k_d^2 \,
\brm{k}_\perp^l}{[B\tau + Ak_d^2 + B \brm{k}_\perp^4]^n} \, .
\end{eqnarray}
This integral can be calculated by the same means as
$M^{(1)}_{ln}$. We obtain the result
\begin{eqnarray}
\label{M25}
&& M^{(2)}_{ln} = \frac{A^{-3/2} \, B^{-(2n-3)/2} \, \Gamma
(\frac{2+l-\varepsilon}{4})}{4 \, (4\pi)^{d/2} \, \Gamma (n) \,
\Gamma (\frac{2-\varepsilon}{2})}
\nonumber \\
&& \times \, \Gamma \left( \frac{4n -l-8 +\varepsilon}{4} \right)
\tau^{- \frac{4n -l-8 +\varepsilon}{4}} \, .
\end{eqnarray}


\begin{thebibliography}{}
\bibitem{footnote1}
Here and in the following we adopt Einstein's summation
convention.
\bibitem{footnote2}
Provided, of course, there are no alligning fields.
\bibitem{footnoteKb}
For convenience, we set the Boltzmann constant equal to one.
\bibitem{footnoteWilson}
The $|_0$ indicates that bare quantities are kept fixed while
taking the derivatives.
\bibitem{footnoteShearModuli}
In our notation these shear moduli are given by $\mu_L = (\omega +
f + 2/d_\perp - 2g)/4$ and $\mu =1$.
\end{thebibliography}
\end{document}